\documentclass[12pt]{article}
\usepackage{epsfig}
\newcommand{\pfrac}[2]{\left(\frac{#1}{#2}\right)}
\newcommand{\real}{\mathop{\rm Re}\nolimits}
\newcommand{\imag}{\mathop{\rm Im}\nolimits}
\newcommand{\ttilde}{\rlap{\raise9pt\hbox{$\scriptstyle\approx$}}}
\begin{document}
\begin{flushright}
MZ-TH/12-35\\
August 2012
\end{flushright}

\begin{center}
{\Large\bf Renormalization group approach to chaotic strings} \\[1truecm]
{\large Stefan Groote$^{1,2}$, Hardi Veerm\"ae$^2$
  and Christian Beck$^3$}\\[.7cm]

$^1$ Institut f\"ur Physik der Johannes-Gutenberg-Universit\"at,\\
  Staudinger Weg 7, 55099 Mainz, Germany\\[.5truecm]
$^2$ F\"u\"usika Instituut, Tartu \"Ulikool,
  T\"ahe 4, 51010 Tartu, Estonia\\[.5truecm]
$^3$ School of Mathematical Sciences,\\
  Queen Mary University of London, Mile End Road, London E1 4NS, UK\\[.5truecm]
\vspace{1truecm}
\end{center}

\begin{abstract}
Coupled map lattices of weakly coupled Chebychev maps, so-called chaotic
strings, may have a profound physical meaning in terms of dynamical models of
vacuum fluctuations in stochastically quantized field theories. Here we
present analytic results for the invariant density of chaotic strings, as well
as for the coupling parameter dependence of given observables of the chaotic
string such as the vacuum expectation value. A highly nontrivial and
selfsimilar parameter dependence is found, produced by perturbative and
nonperturbative effects, for which we develop a mathematical description in
terms of suitable scaling functions. Our analytic results are in good
agreement with numerical simulations of the chaotic dynamics.

\end{abstract}

\newpage

\section{Introduction}
Coupled map lattices (CMLs), as introduced in seminal papers by Kaneko and
Kapral some 29 years ago \cite{kaneko,kapral}, are known to exhibit a rich
structure of complex dynamical phenomena~\cite{kanekobook,Beck:2002,%
Bunimovich:1988,amritkar,carretero}. Of particular interest are CMLs that
consist of locally chaotic maps. For hyperbolic local maps and very small
coupling it can be proved \cite{bunimovich,baladi,Bricmont:1996,%
Jarvenpaa:2001} that a smooth invariant density of the entire CML exists and
that there is ergodic behaviour. However, the case of nonhyperbolic local
maps, e.g.\ of local 1-dimensional maps for which the slope is equal to zero
at some point, is much more complicated from a mathematical point of view, and
much less analytic results are known~\cite{daido,Lemaitre:1998,Lemaitre:1999,%
ding,ruffo,mackey,beck}. This is the realm of chaotic strings.

Chaotic strings are 1-dimensional coupled map lattices of diffusively coupled
Chebychev maps~\cite{beck}. They are intrinsically non-hyperbolic, since the
local maps have an extremum with vanishing slope. They play a very important
role in generalized statistical mechanics models of vacuum
fluctuations~\cite{Beck:2002,Groote:2006,Groote:2007,becknew}, replacing the
noise in stochastically quantized field theories by a deterministic chaotic
dynamics~\cite{parisi,damgaard,Beck:1995}. These type of chaotic theories
are of utmost interest in extensions of the standard model where one
constructs additional sectors of rapidly fluctuating chaotic fields. One
possible physical embedding is to associate the vacum energy generated by the
chaotic strings with dark energy~\cite{dark}. These novel strongly chaotic
models of vacuum fluctuations can help to understand reasons why certain mass
and coupling parameters are realized in nature, others not, and are thus
highly important from a fundamental physics point of view (see
Ref.~\cite{Beck:2002} for more details).

In this paper, however, we will not deal with the above mentioned physical
applications of chaotic strings, but merely concentrate onto the mathematics
of these nonhyperbolic coupled map lattices. There is a highly interesting
selfsimilar dependence of the invariant density on the coupling constant $a$
of the CML, which can be understood by analytic means and which will be the
main subject of this paper.

Let us first recall what has been done so far. In
Refs.~\cite{Groote:2006,Groote:2007} weakly coupled $N$-th order Chebyshev
maps~\cite{beck,Beck:1995,dark,dettmann1,dettmann2} were studied and it was
shown that certain scaling properties with respect to the coupling $a$ can be
calculated perturbatively. However, the fine structure of the parameter
dependence of important observables of the chaotic string, such as the self
energy or vacuum expectation value, could not be explained by these
perturbative methods. In Ref.~\cite{Maher:2008} is was numerically illustrated
that the fine structure does occur only for CMLs of dimension $1$, and not on
lattices of higher dimension. In Ref.~\cite{Groote:2009} short chaotic
strings, consisting only of a small number of lattice points, were analyzed
and it was shown that some of these short strings (with lengths between 3 and
5 lattice sites) show the same fine structure as long strings of the order of
10\,000 sites, while for other short strings the fine structure had a totally
different shape.

In this paper we report on significant progress to better understand the
scaling structure of chaotic strings. We will show that the invariant density
can not only be understood near the edges of the interval on which the local
maps are defined, but in the entire interval region. We will study the
coupling constant dependence of the expectation value of the chaotic field
variable. This expectation is expressed as a particular integral, which is
trivial for the uncoupled case (coupling $a=0$) but exhibits a highly
nontrivial parameter dependence for $a\ne 0$. There are certain invariance
properties of observables of the chaotic string under suitable rescaling of
the coupling that we will describe in terms of a kind of renormalization
group theory. For this purpose we introduce two basic scaling functions, which
we call `draft function' and `blunt function' which describe how parameter
changes of the coupling are generating a selfsimilar pattern and how certain
`excitations' (travelling maxima) in the invariant density are induced. We
also further develop the perturbation theory, and illustrate that there are
perturbative and nonperturbative effects for the chaotic string, with certain
analogy to what is known in quantum field theories.

\vspace{7pt}
This paper is organized as follows. In Sec.~2 we globalize the perturbative
methods developed in Refs.~\cite{Groote:2006,Groote:2007} and apply them to
the vacuum expectation value $\langle\phi\rangle$ of the chaotic field
variable. In Sec.~3 we analyse the integrand leading to $\langle\phi\rangle$
and introduce draft and blunt functions which can help to analytically
describe the self-similar parameter dependence and invariance properties of
the chatic string. In Sec.~4 we discuss simplified models of the scaling
structure which are connected to the perturbation theory. In Sec.~5 we look at
limiting cases for the partition integrals that occur when the vacuum
expectation is calculated. In Sec.~6 we provide analytical expressions for the
(exponential-type) draft function. Finally, in Sec.~7 we draw our conclusions.
More details on explicit calculations can be found in the Appendices.

\section{The perturbative method}
\begin{figure}\begin{center}
\epsfig{figure=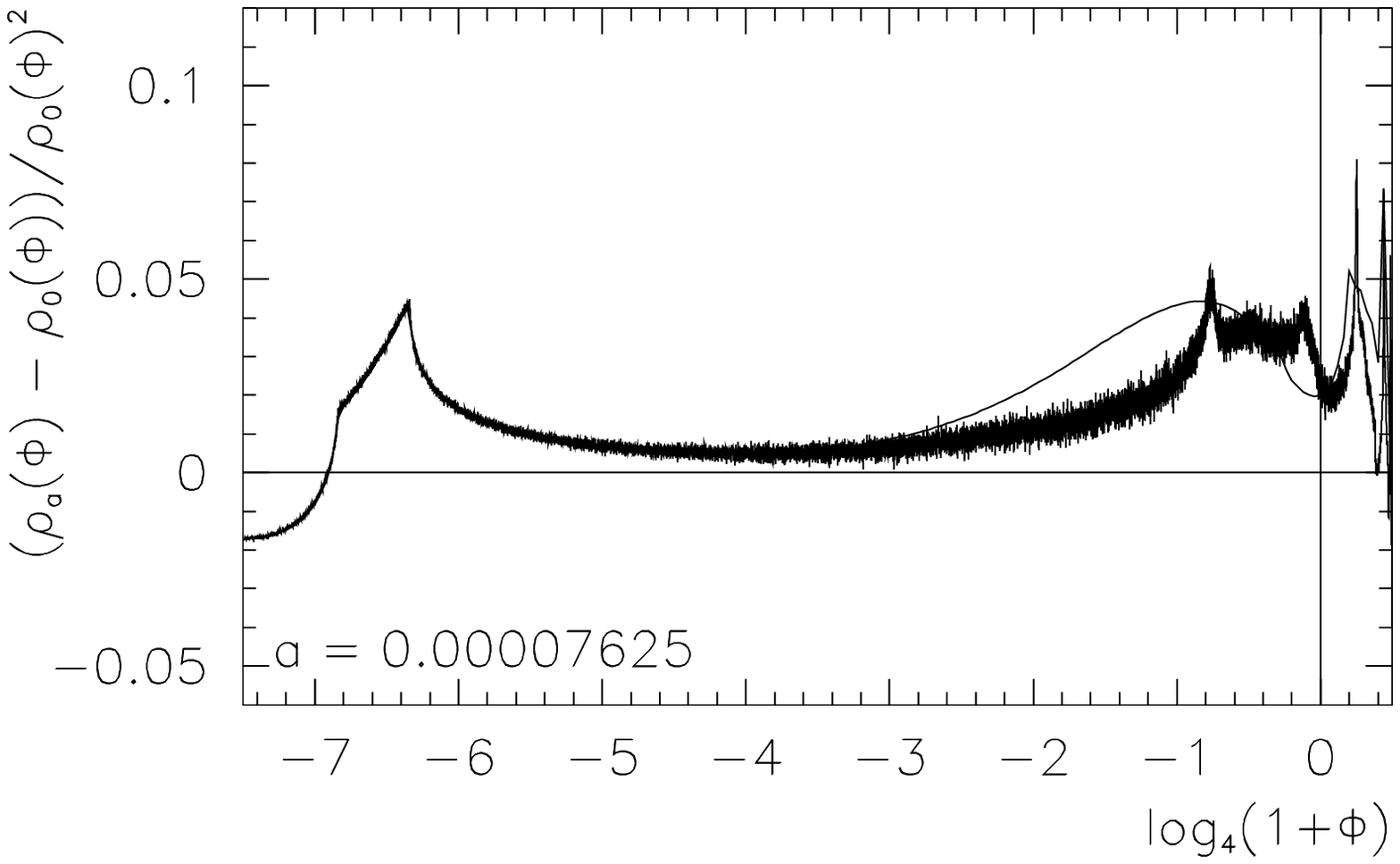, scale=0.8}
\caption{\label{sum237rb}Numerical result obtained by (a) long-term iteration
of the CML and (b) the analytic approximation~(\ref{approx}) at maximal order
$p_{\rm max}=7$. The figure shows the rescaled distribution function
$(\rho_a(\phi)-\rho_0(\phi))/\rho_0^2(\phi)$ at coupling $a=0.00007625$ close to
the lower boundary $\phi=-1$ in dependence on $\log_4(1+\phi)$.}
\vspace{36pt}
\epsfig{figure=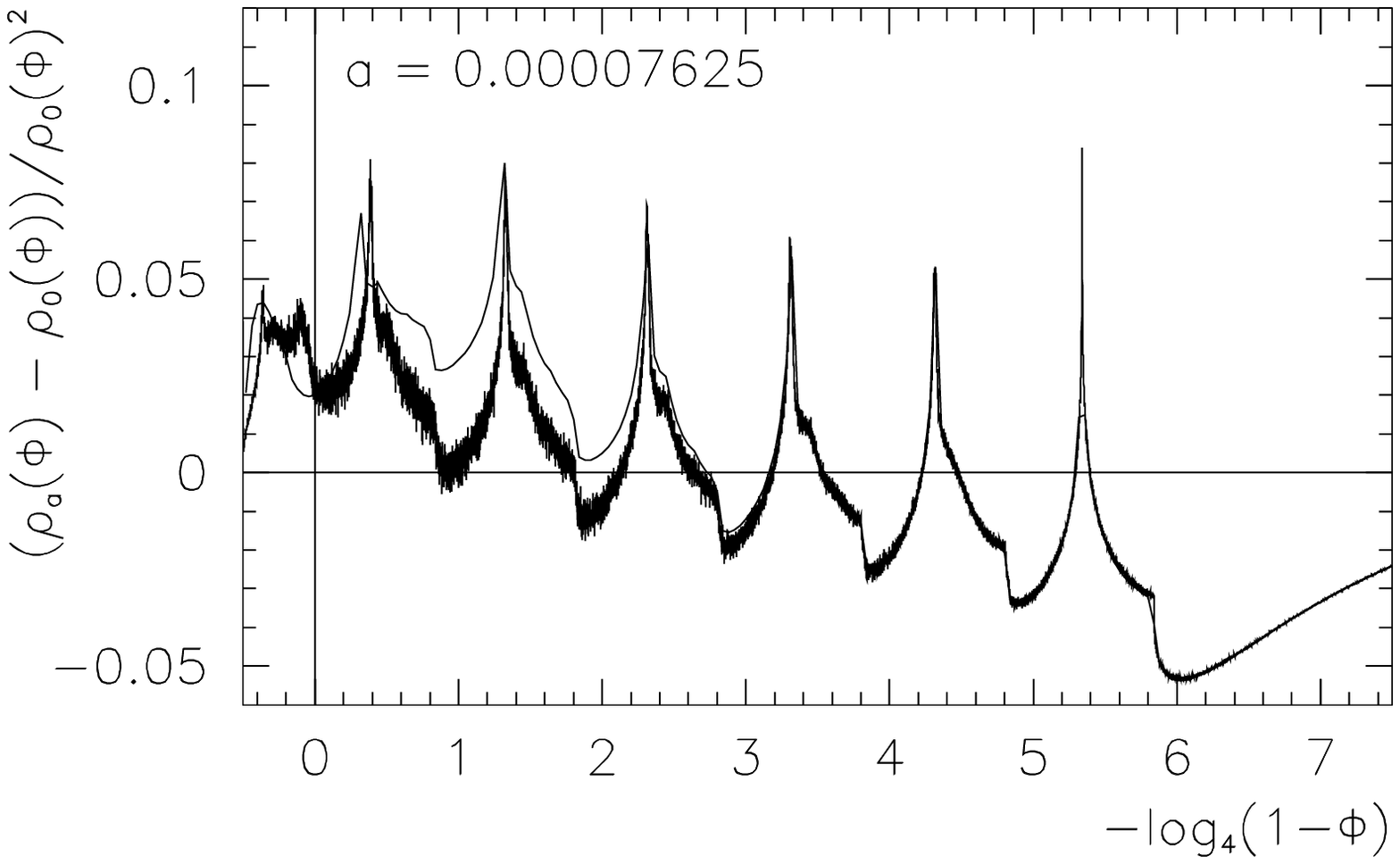, scale=0.8}
\caption{\label{sum237rd} Same as Fig.~1 but close to the upper boundary
$\phi=+1$ in dependence on $-\log_4(1-\phi)$.}
\end{center}\end{figure}

As shown in Ref.~\cite{Groote:2007} in detail, the scaling behaviour of the
self energy (and interaction energy) of the chaotic string based on Chebyshev
maps $T_N(\phi)$ of order $N$ is directly related to the scaling behaviour of
the distribution function (the 1-point invariant density) $\rho(\phi)$ of the
string. For the chaotic string of type 2B (cf.\ Ref.~\cite{Beck:2002}) with
iterative prescription
\begin{equation}\label{iter2B}
\phi^{n+1}_i=(1-a)T_2(\phi^n_i)+\frac a2(\phi^n_{i+1}+\phi^n_{i-1})
  =:T_{2a}(\phi^n_i;\phi^n_{i+1}+\phi^n_{i-1})
\end{equation}
the distribution function close to the upper boundary $\phi=1$ turned out to
be a sum $\rho_a(1-ax)\approx\sum_{p=1}^{p_{\rm max}}\rho_a^{(p)}(1-ax)$ of functions
$\rho_a^{(p)}$ (in the following sometimes called $p$-iterates)
with\footnote{Compared to Ref.~\cite{Groote:2007} we use a different
definition~(\ref{defrNp}) for the function $r_N^p(\phi)$ in order to keep it
positive. Note that by convention the integration range is given by the
positivity condition for the radicand~\cite{Groote:2007}.}
\begin{equation}\label{approxp}
\rho_a^{(p)}(1-ax)=\frac1{\pi\sqrt{2a(1-a)}}\int\frac{\rho_0(\phi_+)d\phi_+
  \rho_0(\phi_-)d\phi_-}{4^p\sqrt{x/4^p-r_2^p(\phi_+)-r_2^p(\phi_-)}}
\end{equation}
while close to the lower boundary $\phi=-1$ we obtain
$\rho_a(ay-1)\approx\rho_a^{(0)}(ay-1)$ with
\begin{equation}\label{approxm}
\rho_a^{(0)}(ay-1)=\frac1{\pi\sqrt{2a(1-a)}}\int\frac{\rho_0(\phi_+)d\phi_+
  \rho_0(\phi_-)d\phi_-}{\sqrt{y-(\phi_++\phi_-)/2-1}}.
\end{equation}
$\rho_0(\phi)=1/\pi\sqrt{1-\phi^2}$ is the invariant density for the uncoupled
string ($a=0$). We use the notation $\phi_+:=\phi^n_{i+1}$, $\phi_-:=\phi^n_{i-1}$,
$\phi:=\phi^n_i$, and
\begin{equation}\label{defrNp}
r_N^p(\phi):=\frac12\sum_{q=0}^p\frac{1-T_{N^q}(\phi)}{N^{2q}}.
\end{equation}
Both approximations can be formally combined to the formula
\begin{equation}\label{approx}
\rho_a(\phi)=\sum_{p=0}^{p_{\rm max}}\rho_a^{(p)}(\phi)
\end{equation}
This approximation is valid only close to the boundaries $\phi =\pm 1$. This
is shown in Fig.~\ref{sum237rb} (close to the lower boundary) and
Fig.~\ref{sum237rd} (close to the upper boundary). For the main part of the
interval $\phi\in[-1,1]$, however, there are large deviations between numerical
results obtained by iteration of the CML and the above formula, regardless of
the order $p_{max}$ of the approximation. Nevertheless, a main result of this
paper is an extension to a formula that is valid for general values of $\phi$.

\subsection{Extension to the whole range}
The key observation for this extension is given by the insight that the
integrand of
\begin{equation}
\rho_a^{(0)}(\phi)=\frac1{\pi\sqrt{1-a}}\int\frac{\rho_0(\phi_+)d\phi_+
  \rho_0(\phi_-)d\phi_-}{\sqrt{2(\phi+1)-a(\phi_++\phi_-+2)}}
\end{equation}
is given by the derivative of the inverse map of the map~(\ref{iter2B}),
\begin{equation}\label{inviter2B}
\frac1{2\sqrt{1-a}\sqrt{2(\phi+1)-a(\phi_++\phi_-+2)}}
  =\frac{d}{d\phi}T_{2a}^{-1}(\phi;\phi_++\phi_-)
  =:T_{2a}^{-1\prime}(\phi;\phi_++\phi_-).
\end{equation}
Of course, the inverse $T_{2a}^{-1}(\phi;\phi_++\phi_-)$ of the function
$T_{2a}(\phi;\phi_++\phi_-)$ given by the quadratic equation~(\ref{iter2B})
with $T_2(\phi)=2\phi^2-1$ is not unique. We have used the positive square root
$T_{2a}^+(\phi;\phi_++\phi_-)$ according to the notation
\begin{equation}
T_{2a}^\pm(\phi;\phi_++\phi_-)
  :=\pm\sqrt{\frac{2(\phi+1)-a(\phi_++\phi_-+2)}{4(1-a)}}
  =T_{2a}^{-1}(\phi;\phi_++\phi_-)
\end{equation}
in order to write
\begin{equation}
\rho_a^{(0)}(\phi)=\frac2\pi\int\rho_0(\phi_+)d\phi_+\rho_0(\phi_-)d\phi_-
  \frac{d}{d\phi}T_{2a}^+(\phi;\phi_++\phi_-).
\end{equation}
Proceeding to the first iterate, we extend to
\begin{eqnarray}\label{rhoa1ext}
\rho_a^{(1)}(\phi)&=&-\frac2\pi\int\rho_0(\phi_+)d\phi_+\rho_0(\phi_-)d\phi_-
  \times\strut\nonumber\\&&\strut\qquad\times
  \frac{d}{d\phi}T_{2a}^+\left(T_{2a}^-(\phi;T_2(\phi_+)+T_2(\phi_-));
  \phi_++\phi_-\right).
\end{eqnarray}
For details of this and the following see Appendix~A. For an arbitrary order
$p$ we finally arrive at
\begin{eqnarray}
\lefteqn{\rho_a^{(p)}(\phi)\ =\ -\frac2\pi\int\rho_0(\phi_+)d\phi_+
  \rho_0(\phi_-)d\phi_-\times\strut}\\&&\strut\times
  \frac d{d\phi}T_{2a}^+\left(T_{2a}^+\left(\ldots
  T_{2a}^-\left(\phi;T_{2^p}(\phi_+)+T_{2^p}(\phi_-)\right)\ldots;
  T_2(\phi_+)+T_2(\phi_-)\right);\phi_++\phi_-\right).\nonumber
\end{eqnarray}

\subsection{The ``path of roots''}
The path ``$+\cdots+-$'' for selecting the nested square roots turns out to
lead to an excellent approximation close to the upper limit $\phi=+1$. Of
course we have to ask why this path needs to be taken and not a different one.
The answer to this question is related to the trajectory that a starting value
close to zero takes under nearly unperturbed iterations. Starting from
$\phi=0$ we obtain $T_2(0)=-1$, $T_2(T_2(0))=+1$, $T_2(T_2(T_2(0)))=+1$ and so
forth. This is just the time-reverse of the relevant ``path of roots''.
However, if we want to extend the approximation of the invariant density in
the way mentioned above, we need to take into account all paths. It turns out
that a great amount of other paths returns contributions of lower order.
Moreover, these contributions add up in a way that a general factor $2/\pi$ is
factored out. The final result (understood as sum over all paths of roots)
\begin{eqnarray}\label{pathroot}
\lefteqn{\rho_a^{(p)}(\phi)\ =\ \int\rho_0(\phi_+)d\phi_+\rho_0(\phi_-)d\phi_-
  \times\strut}\\&&\strut
  \frac d{d\phi}T_{2a}^{-1}\left(T_{2a}^{-1}\left(T_{2a}^{-1}\left(\ldots;
  (\phi;\ldots)\ldots;T_2(T_2(\phi_+))+T_2(T_2(\phi_-))\right);
  T_2(\phi_+)+T_2(\phi_-)\right);\phi_++\phi_-\right)\nonumber
\end{eqnarray}
for {\em a single\/} $p$-iterate (without adding lower iterates) approximates
the distribution function to an increasing degree with increasing order $p$.
More details on the derivation of the above equation can be found in
Appendix~B.

\begin{figure}[ht]\begin{center}
\epsfig{figure=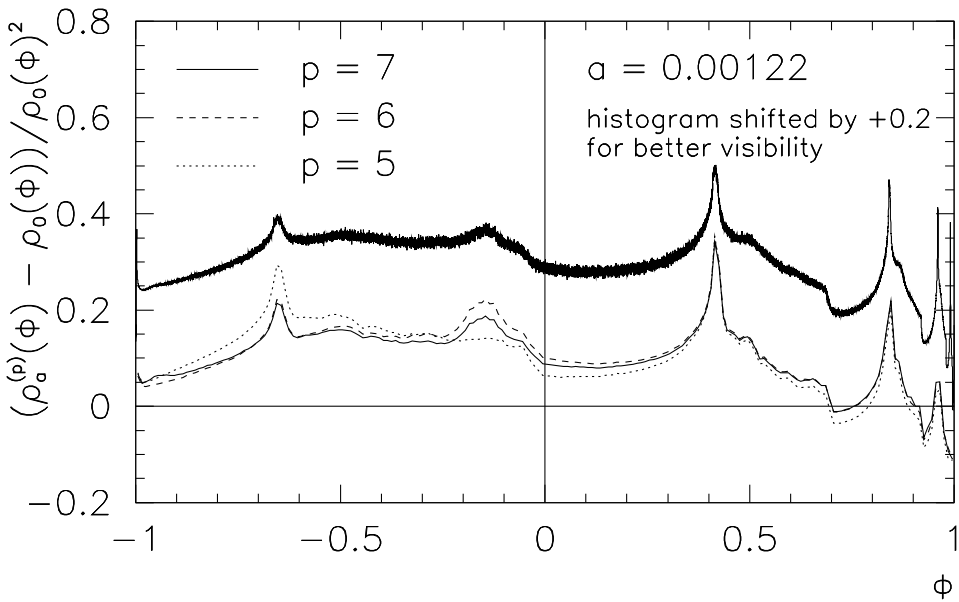, scale=0.8}
\caption{\label{sum217w}Comparison of the rescaled $\rho_a^{(p)}(\phi)$
($p=5,6,7$) with the numerically obtained histogram result. In this plot the
histogram is shifted by the amount $+0.2$ for better visibility.}
\end{center}\end{figure}

\vspace{7pt}
In Fig.~\ref{sum217w} the coincidence of higher $p$-iterates with the
distribution function is shown. For the coupling we have chosen a rather high
value $a=0.00122$ as we are no longer interested in the perturbative region
close to the boundary but in nonperturbative effects in the middle of the
interval. As compared to Figs.~1 and~\ref{sum237rd}, the position and magnitude
of the local maxima (often called `excitations' in the following) are mirrored
in a precise way. Even the excitation close to $\phi=-0.15$, i.e.\ in a highly
nonperturbative region, is exactly modeled.

\vspace{7pt}
The excitation at $\phi\approx -0.15$ is special. Indeed, if we look at a
sequence of diagrams similar to Fig.~3 where the coupling $a$ gradually
decreases, then this excitation moves to the left while all others move to the
right. These and other counter movements account for the so-called 'fine
structure' of observables of the chaotic string. By this we mean a nontrivial
local $a$-dependence of observables as illustrated e.g.\ in Fig.~4 and 5. The
reason for the different behaviour of the above excitation is that this
excitation belongs to a different ``path of roots'' than the ordinary one
(``$+\cdots+-$''). It can be associated with a process that is mirrored at the
lower boundary. This becomes apparent when we look at a sequence of diagrams
as in Fig.~3 again. At the moment when the extraordinary excitation disappears
at $\phi=-1$, a new ordinary excitation appears at this point and moves to the
right.

\begin{figure}\begin{center}
\epsfig{figure=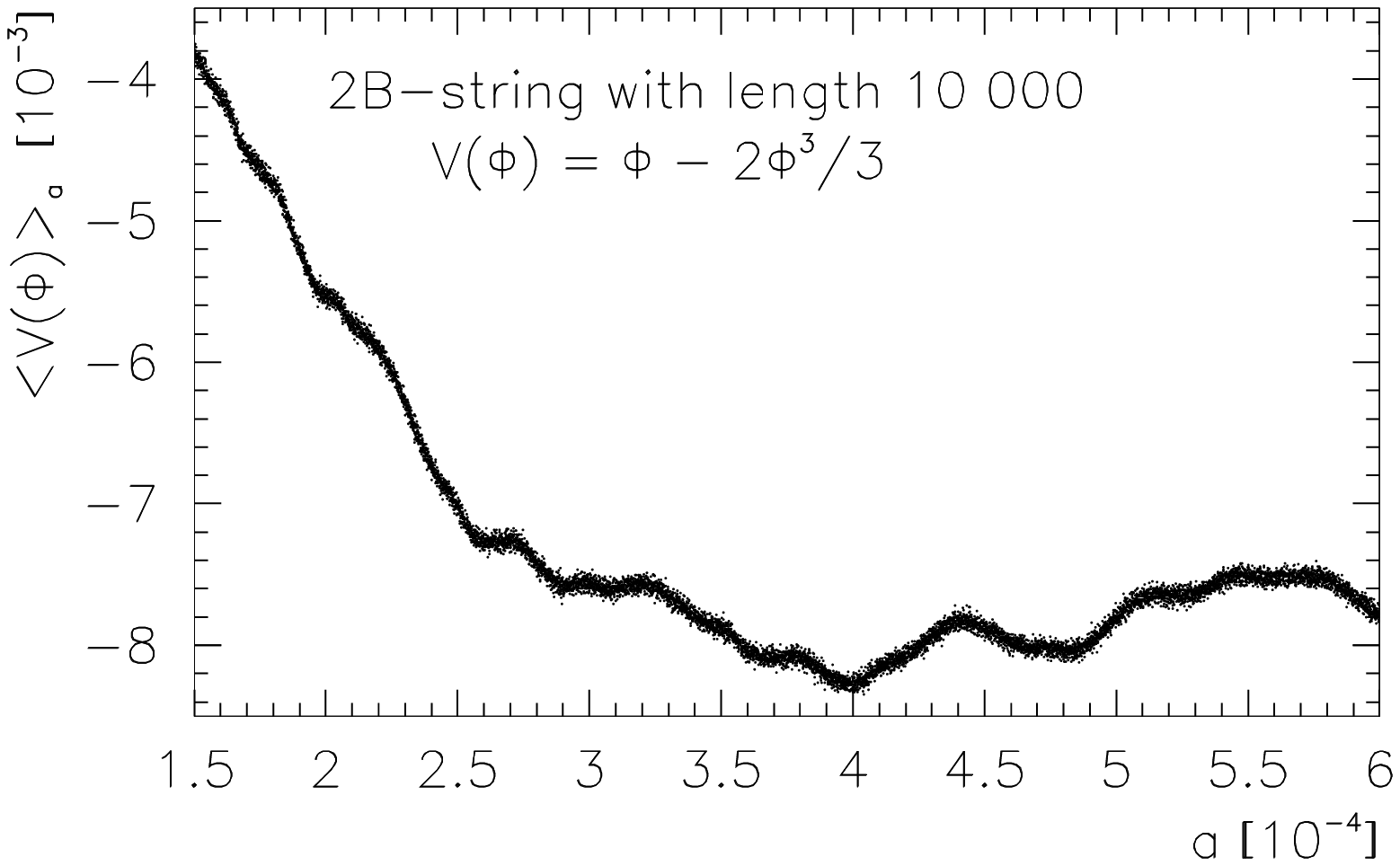, scale=0.8}
\caption{\label{self2bs}Expectation value of the self energy $V(\phi)$ in the
range $a\in[0.00015,0.00060]$ for a 2B-string of 10\,000 lattice points and
10\,000 iterations}
\vspace{12pt}
\epsfig{figure=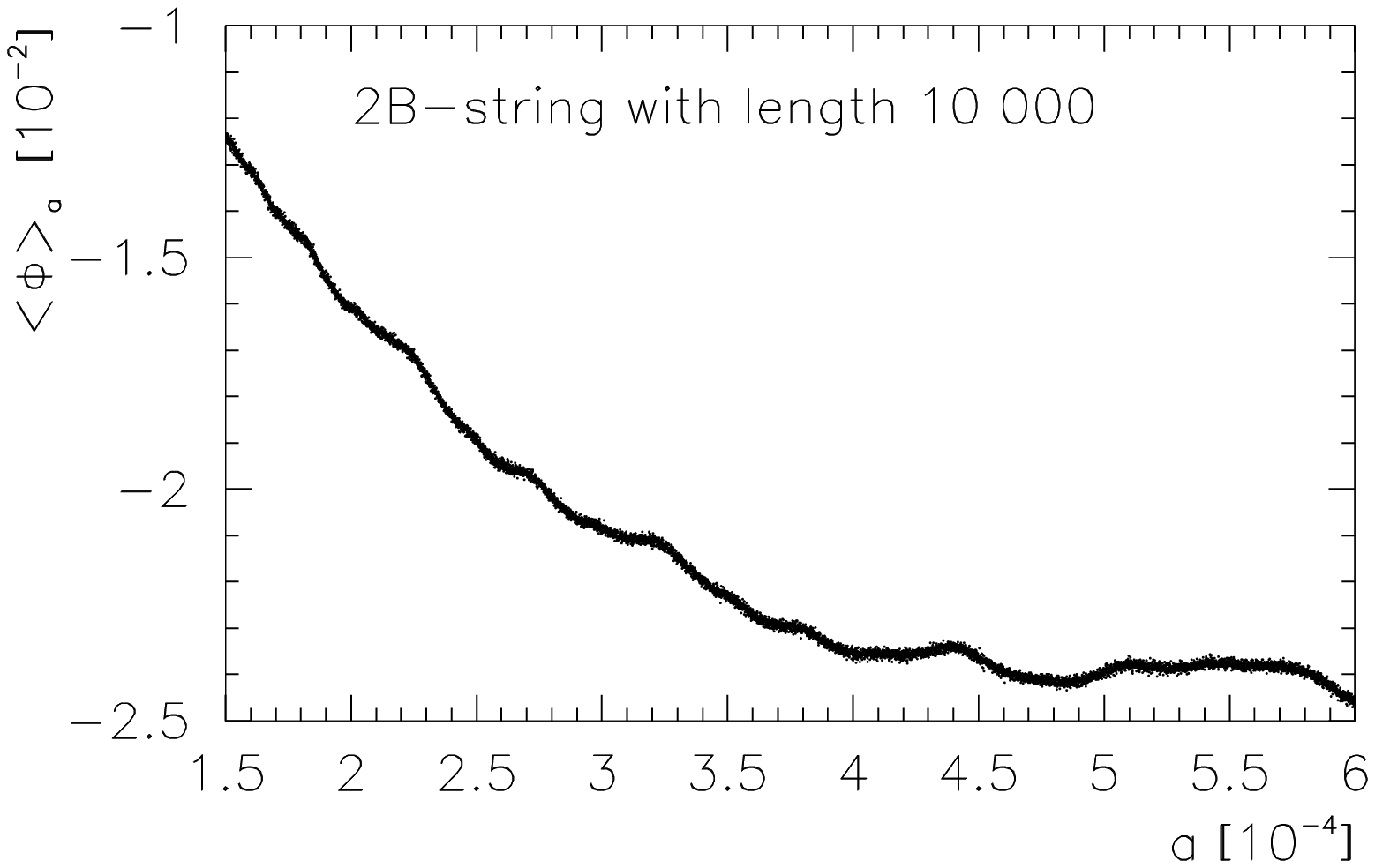, scale=0.8}
\caption{\label{self2b1}Expectation value of $\phi$ in the range
$a\in[0.00015,0.00060]$ for a 2B-string of 10\,000 lattice points and
10\,000 iterations}
\end{center}\end{figure}

\subsection{Expectations of observables}
\begin{figure}\begin{center}
\epsfig{figure=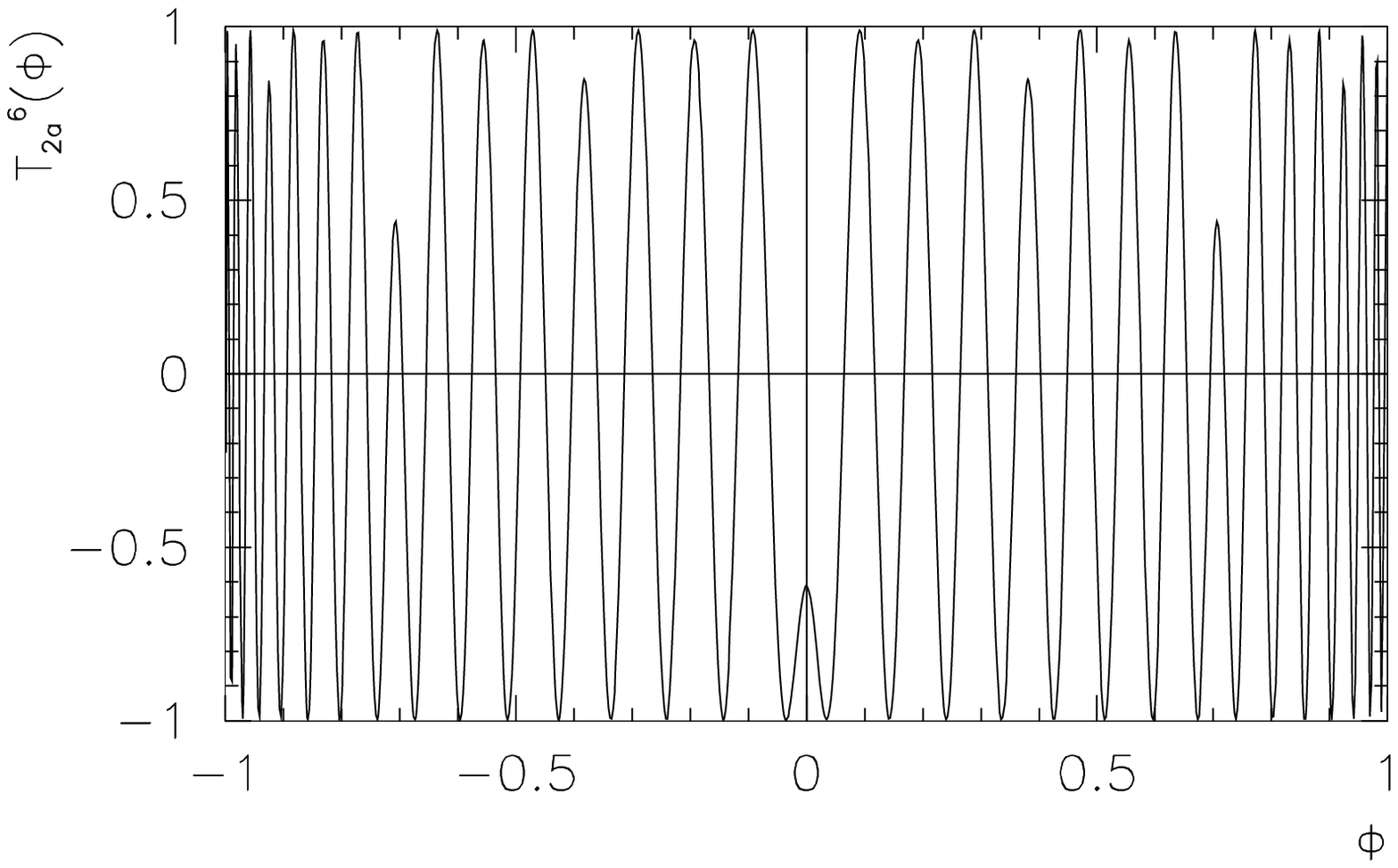, scale=0.8}
\caption{\label{intgran6}Integrand of Eq.~(\ref{expectphi}) for $p=5$ at
  $a=0.00122$ and $\phi_-=\phi_+=1$}
\end{center}\end{figure}

The fine structure we are trying to understand in this paper is the
$a$-dependence of observables (functions of $\phi$) such as e.g.\ the self
energy $V(\phi)=\phi-2\phi^3/2$ calculated with respect to the distribution
function $\rho_a (\phi)$ (cf.\ Fig.~\ref{self2bs}). These types of
expectations of observables as a function of $a$ turn out to be of utmost
interest for the physical applications described in Ref.~\cite{Beck:2002}.
It turns out that the vacuum expectation value $\langle\phi\rangle$ of the
string leads to the same (but less pronounced) fine structure, as it is shown
in Fig.~\ref{self2b1}. Therefore, in the following we will concentrate onto
the calculation of $\langle\phi\rangle=\int\phi\rho_a (\phi)$. Approximating
$\rho_a (\phi)$ by the $p$-iterate $\rho_a^{(p)}(\phi)$, one can perform a
$(p+1)$-fold integration by parts to obtain
\begin{eqnarray}\label{expectphi}
\lefteqn{\langle\phi\rangle_a\ =\ \int\phi\rho_a^{(p)}(\phi)d\phi
  \ =\ \int\rho_0(\phi_+)d\phi_+\rho_0(\phi_-)d\phi_-T_{2a}
  \Big(T_{2a}\Big(T_{2a}\Big(\ldots T_{2a}\left(\phi^{(p)};\phi_++\phi_-\right)
  \ldots;}\nonumber\\&&
  T_{2^{p-2}}(\phi_+)+T_{2^{p-2}}(\phi_-)\Big);T_{2^{p-1}}(\phi_+)
  +T_{2^{p-1}}(\phi_-)\Big);T_{2^p}(\phi_+)+T_{2^p}(\phi_-)\Big)d\phi^{(p)}.
  \qquad\qquad
\end{eqnarray}
This equation reminds us of results obtained in Ref.~\cite{Groote:2009} for an
open string of length 3.\footnote{Note that due to a numerical error, in
Ref.~\cite{Groote:2009} is was erroneously stated that the open string of
length~3 does not reproduce the fine structure. This will be corrected in an
erratum to Ref.~\cite{Groote:2009}.}
In order to better understand how the excitation can switch to a different
``path of roots'' if $a$ is changed we take a detailed look at the integrand
in Eq.~(\ref{expectphi}) as a function of $\phi^{(p)}$ (renamed as $\phi$) for
different values of $(\phi_+,\phi_-)$. For $(\phi_+,\phi_-)=(+1,+1)$ the
integrand at $a=0.00122$ is shown in Fig.~\ref{intgran6}.
We see that while the minima of this strongly oscillating function take the
value $-1$, some of the maxima take values less than $1$. The most extreme
situation is found at $\phi=0$ where the maximum is on the way to decay
completely if $a$ changes in a significant way. When this happens, the two
neighbouring minima unite into one new minimum. From the calculations in
Refs.~\cite{Groote:2006,Groote:2007} it can be seen that the height of maxima
is related to excitations of the distribution function. The decay of a maximum
at $\phi=0$ for increasing values of $a$ thus means that an excitation ceases
to exist at the lower boundary. The vanishing of the extraordinary excitation
mentioned before is related to such a formation of a united minimum.

\section{Structural changes of the generating partition}
The change described before is a change of the topological structure because
one of the local oscillations which partitions the interval $[-1,1]$ into
$2^p$ intervals vanishes. Indeed, the different partitions stand in direct
correspondence to the different roots of the integrand. For vanishing coupling
$a=0$ the maxima and minima are located exactly at $\sin(\pi t/2)$ for the
values $t=\pm 2n/2^p$ and $t=\pm(2n+1)/2^p$, respectively, where
$n=0,1,\ldots,2^{p-1}$ counts the extrema. The minima divide the interval
$[-1,1]$ into $2^p$ intervals which we call the elements of the generating
partition (see Fig.~6). If one of these elements vanishes, the topological
structure of the partition changes.

\vspace{7pt}
One can use the generating partition to split up the integral in
Eq.~(\ref{expectphi}) into a sum of $2^p$ partition integrals. From this point
of view, the decay of a maximum means the vanishing of the corresponding
partition integral. In the following we take a closer look on how an
increasing coupling $a$ influences the contribution of the integrand through
the different elements of the generating partition. In order to distribute the
elements of the generating partition uniformly onto the interval $[-1,1]$, we
use a new variable $t$, with $\phi=\sin(\pi t/2)$.

\subsection{Decay of the maxima}
A general feature of the integrand is that maxima which for $a=0$ are located
at equally spaced positions (``starting positions'') $t_0$ decay in a similar
way when $a$ is increased. This property forms the basis for our
renormalization group approach in the following. Let us assign a class $k$ to
certain subsets of maxima as follows: The maxima with starting positions
$t_0=0$ and $\pm 1$ are of class $k=0$, the maxima with starting positions
$t_0=\pm 1/2$ are of class $k=1$, the maxima with starting positions
$t_0=\pm 1/4$, $\pm 3/4$ are of class $k=2$, and so on. In general, the
starting positions of class $k$ are given by $t_0=\pm(2n+1)/2^k$,
$n=0,1,\ldots,2^{k-1}-1$. We note the similarity with the definition of
$k$-cylinders in the thermodynamic formalism of dynamical
systems~\cite{beck-schloegl}. For $(\phi_+,\phi_-)=(+1,+1)$ the height of a
maximum of class $k$ is given by
\begin{equation}\label{blunting}
b=\cos(\sqrt a2^{p-k+1}).
\end{equation}
From Eq.~(\ref{blunting}) we read off that for a specified scaling interval
where $a$ increases by a factor $4$ (e.g.\ $a\in[0.00015,0.00060]$), precisely
one single specified class of maxima (and, therefore, partition integrals)
disappears, and this disappearence takes place synchronously. Apparently the
successive decay of maxima produces global scaling features, such as the
invariance under the transformation $a\to 4a$, but it is not responsible for
the fine structure within a given scaling region, which is more related to
infinitesimal changes of $a$. To better understand the fine structure, in the
following we will introduce two important auxiliary functions: Draft functions
and blunt functions.

\begin{figure}\begin{center}
\epsfig{figure=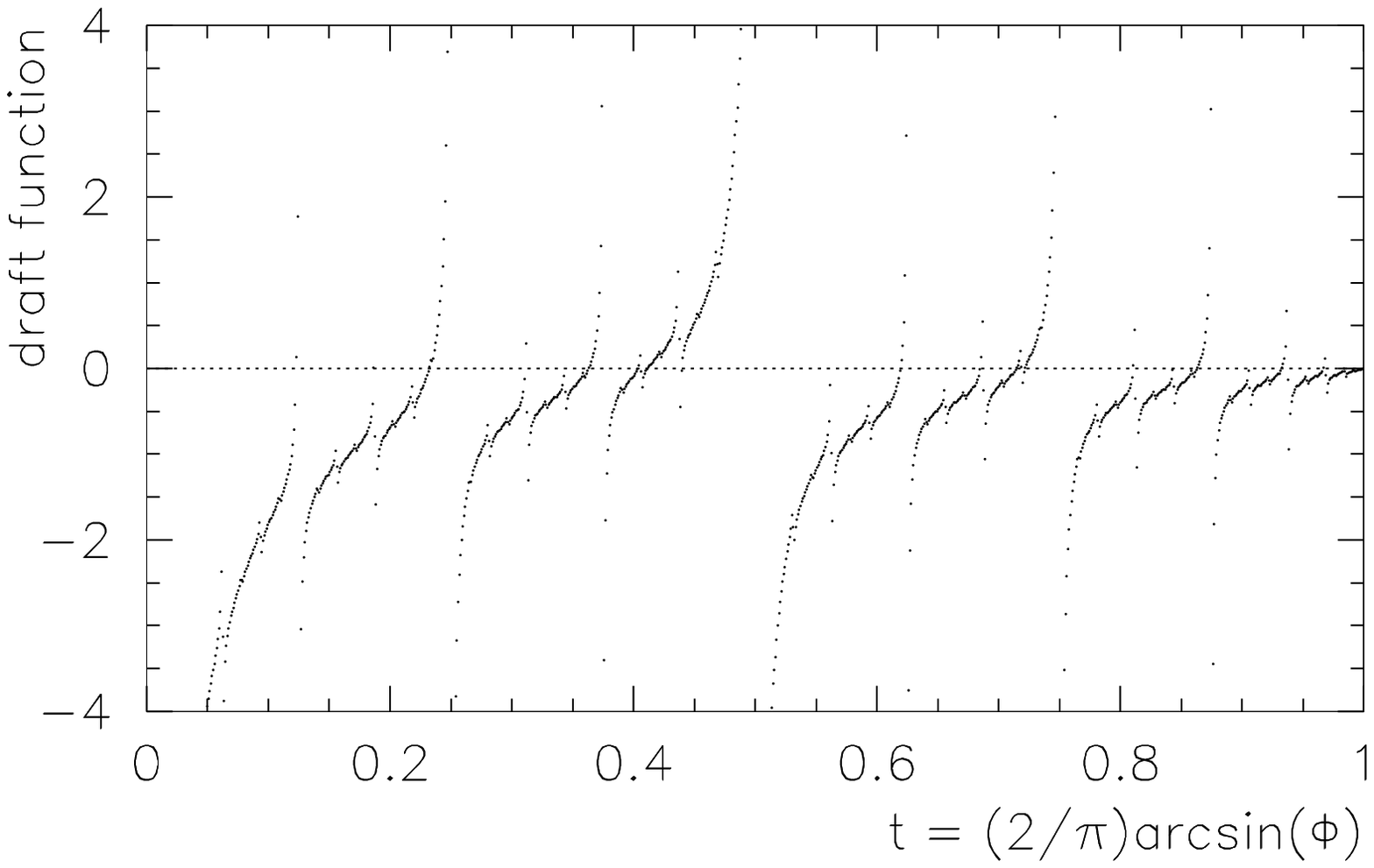,scale=0.6}
\caption{\label{draftpp}Draft function for $\phi_+=\phi_-=+1$}
\vspace{12pt}
\epsfig{figure=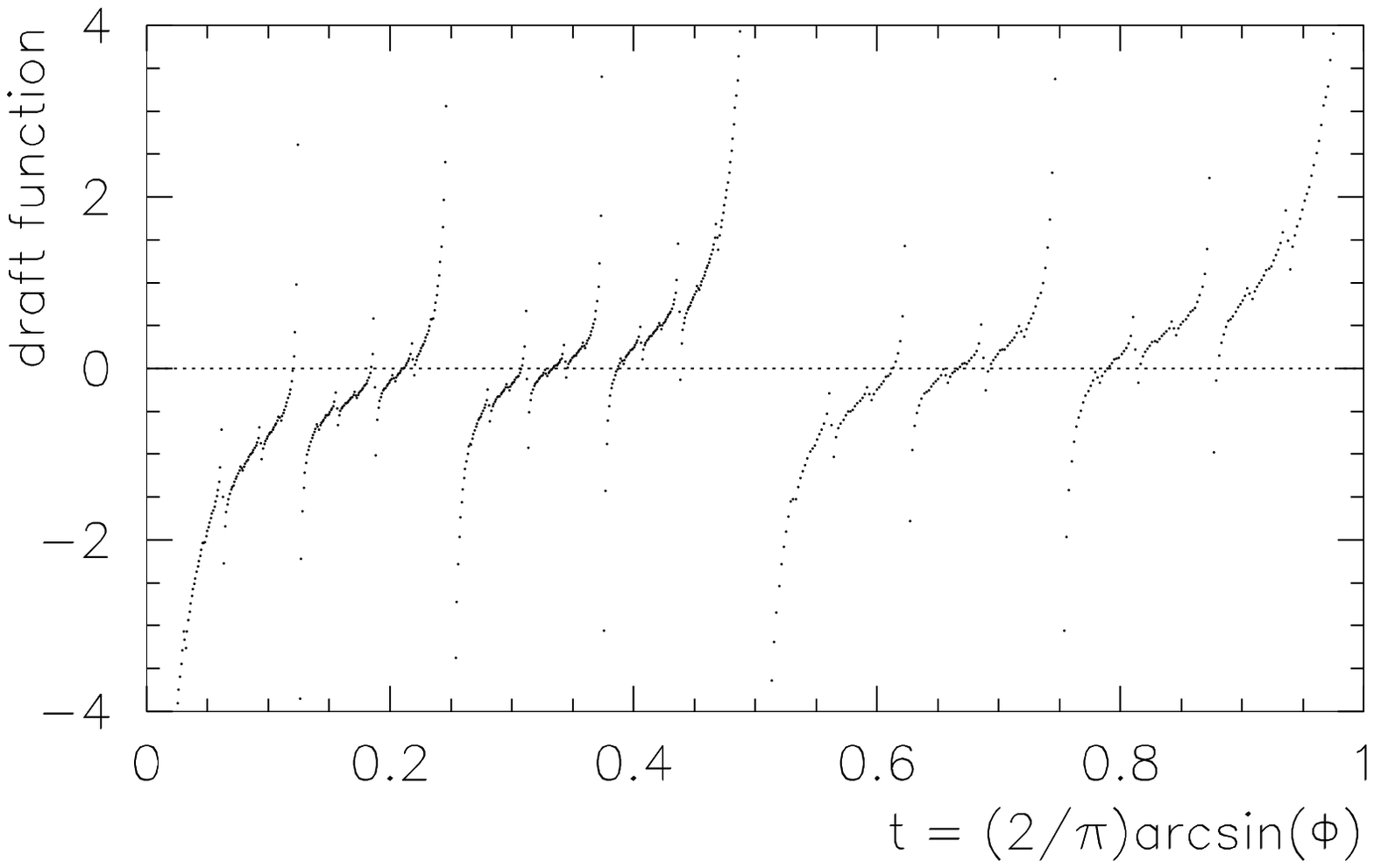,scale=0.6}
\caption{\label{draftpm}Draft function for $\phi_+=+1$ and $\phi_-=-1$}
\vspace{12pt}
\epsfig{figure=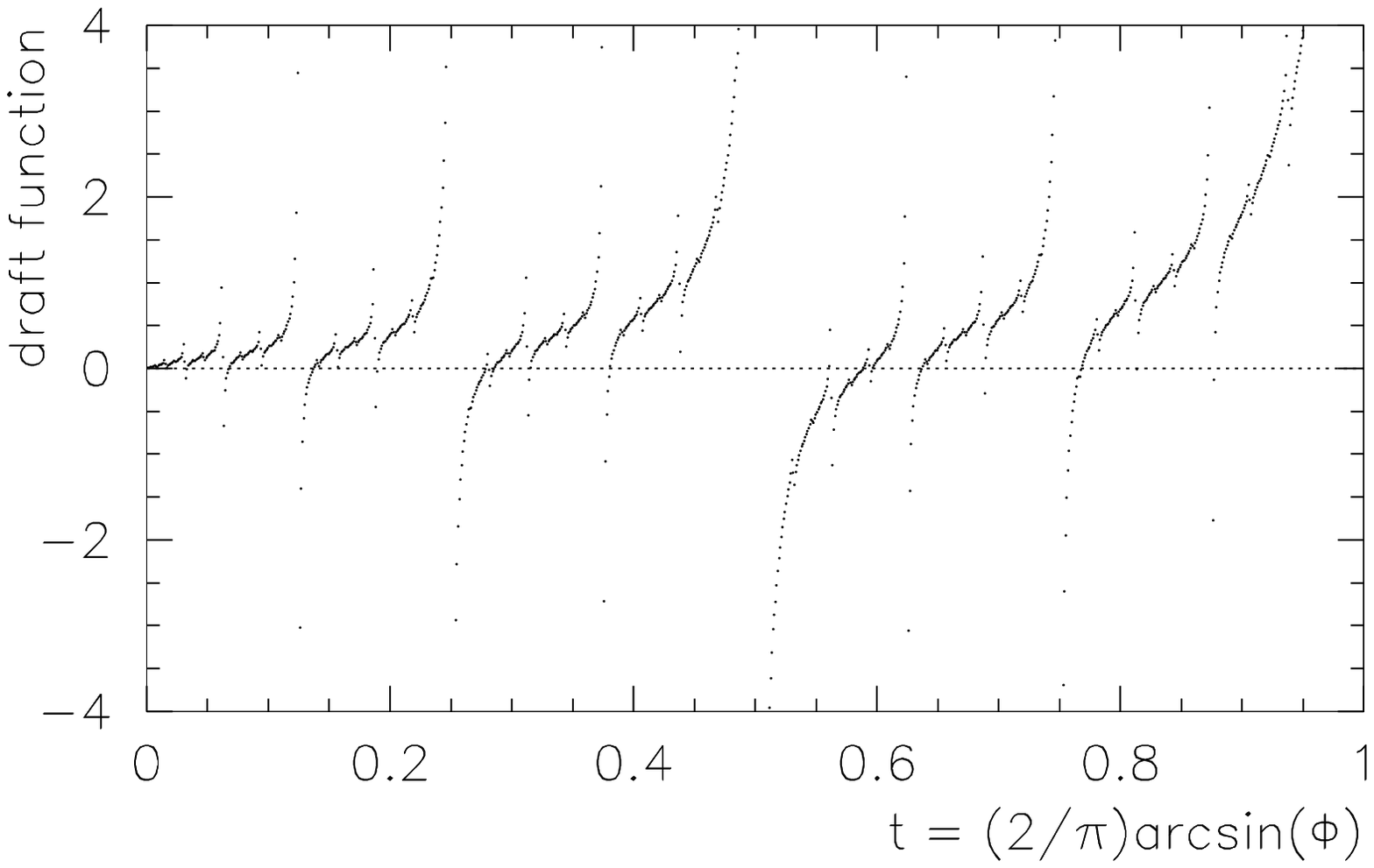,scale=0.6}
\caption{\label{draftmm}Draft function for $\phi_+=\phi_-=-1$}
\end{center}\end{figure}

\subsection{The draft function}
If elements of the generating partition vanish, the neighbouring partition
elements fill the gap, being drawn close to the position where the topological
change happens. The difference between the starting position of a given
maximum and the position for $a\ne 0$ turns out to be a linear function in
$a$. The slope, therefore, is a well-defined function called draft function,
$d(\phi_0)=d\phi_a(a;\phi_0)/da$ where $\phi_a(a;\phi_0)$ is the position of
the maximum with $\phi_a(0;\phi_0)=\phi_0$. For large $p$ and $k$ this is a
fractal function, and the basic features visible in a plot are independent of
$p$ and $k$. The function is shown in Figs.~\ref{draftpp}, \ref{draftpm}
and~\ref{draftmm} for $(\phi_+,\phi_-)=(+1,+1)$, $(+1,-1)$ (which is the same
as for $(-1,+1)$) and $(-1,-1)$. We found an analytical expression for the
draft function. With $\phi_0=\sin(\pi t_0/2)$ the analytical formula for the
draft function of class $k$ reads
\begin{equation}\label{decomp}
\tilde d^{(k)}(t_0;\phi_+,\phi_-)=\frac12\left(\tilde d^{(k)}(t_0;\phi_+)
  +\tilde d^{(k)}(t_0;\phi_-)\right)
\end{equation}
(the tilde indicates the linearization of the first argument by replacing
$\phi_0=\sin(\pi t_0/2)$ by $t_0$) where
\begin{equation}\label{dk1}
\tilde d^{(k)}(t_0;\phi_\pm)=\tilde d^{(k)}(t_0)
  +\sum_{l=0}^{k-1}T_{2^l}(-\phi_\pm)\tilde d_l^{(k)}(t_0)
\end{equation}
and
\begin{equation}
\tilde d^{(k)}(t_0)=2^{-k}d_{-1}^{(k)}(t_0)-\sum_{l=0}^{k-1}
  \left(2^{1-k+l}+\cos(2^l\pi t_0)\right)\tilde d_l^{(k)}(t_0),\qquad
\tilde d_l^{(k)}(t_0)=\frac1{2^l\pi\sin(2^l\pi t_0)}.
\end{equation}
The draft function $\tilde d^{(k)}(t_0;\phi_+,\phi_-)$ is smooth for
$t_0=\pm(2n+1)/2^k$, $n=0,1,\ldots,2^{k-1}-1$ but singular for values $t_0$
corresponding to a class smaller than $k$.

\subsection{The blunt function}
Eq.~(\ref{blunting}) is valid only for the specific value
$(\phi_+,\phi_-)=(+1,+1)$ of the neighbours. For general neighbouring values
$\phi_+, \phi_-$ the height of maxima of class $k$ is given by
\begin{equation}\label{blunt}
b_{p,k}(\phi_+,\phi_-)=\cos\left(2^{p-k+1}\sqrt{ar_2^\infty(-T_{2^k}(\phi_+),
  -T_{2^k}(\phi_-))}\right)
\end{equation}
where $r_N^p$ is defined in Eq.~(\ref{defrNp}). We call $b_{p,k}$ the `blunt
function'. Both draft and blunt functions are important for our
renormalization group description in the following. Basically the draft
function describes how positions of maxima change under parameter changes, and
the blunt function describes how the values of the maxima themselves change.

If the integrand for different values of $(\phi_+,\phi_-)$ is affected by a
relative draft, the blunting is accelerated because after integration over
$\rho_0(\phi_+)\rho_0(\phi_-)$ the integrand (in $\phi$) is smeared out, giving
rise to a premature breakdown of the integral.

\begin{figure}\begin{center}
\epsfig{figure=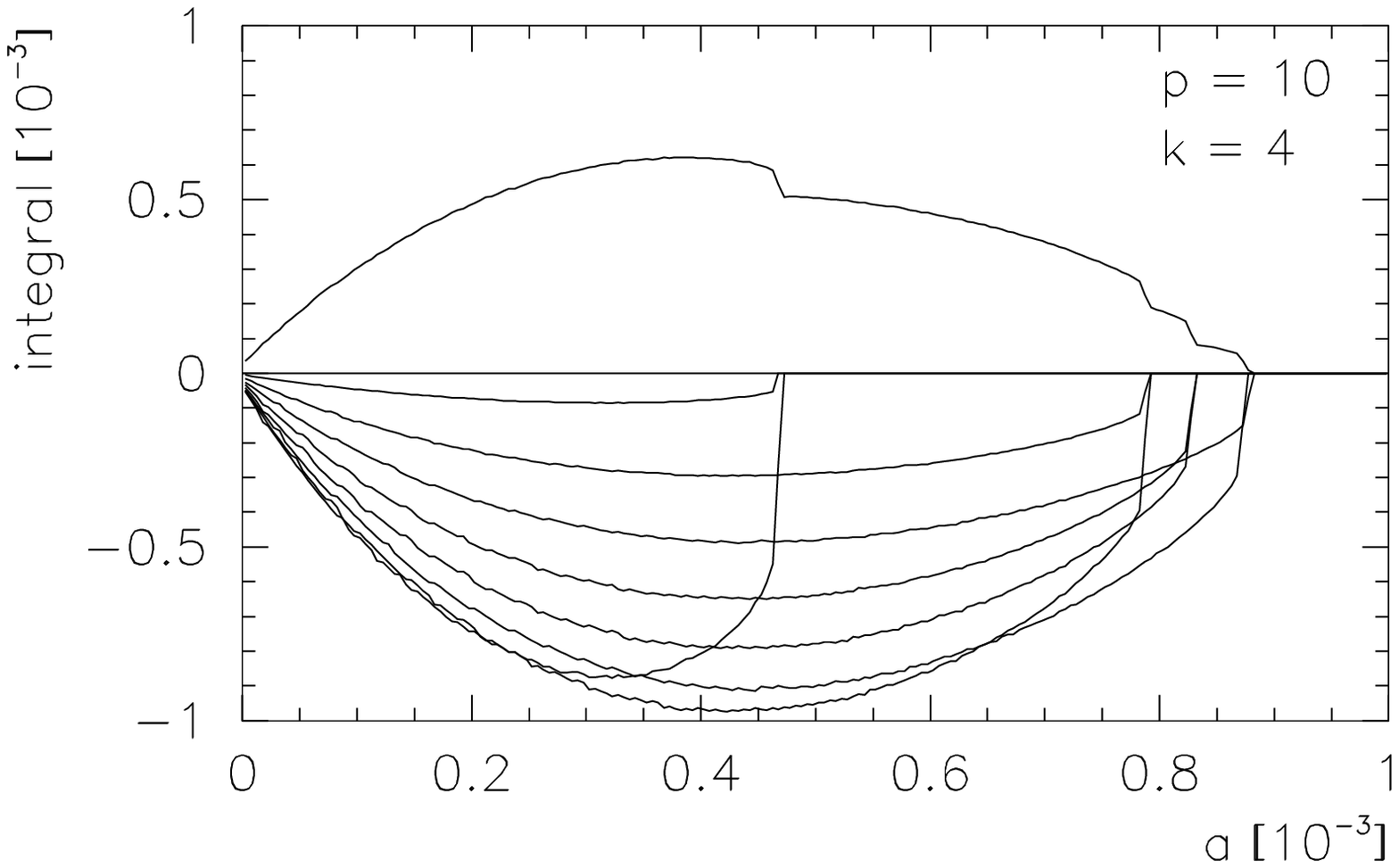, scale=0.8}
\caption{\label{fine2b04a}Contributions and sum of all the partition integrals
for $p=10$ and $k=4$ that vanish in the displayed interval for $a$. For better
visibility, the sum is multiplied by $-2^{1-k}$.}
\vspace{36pt}
\epsfig{figure=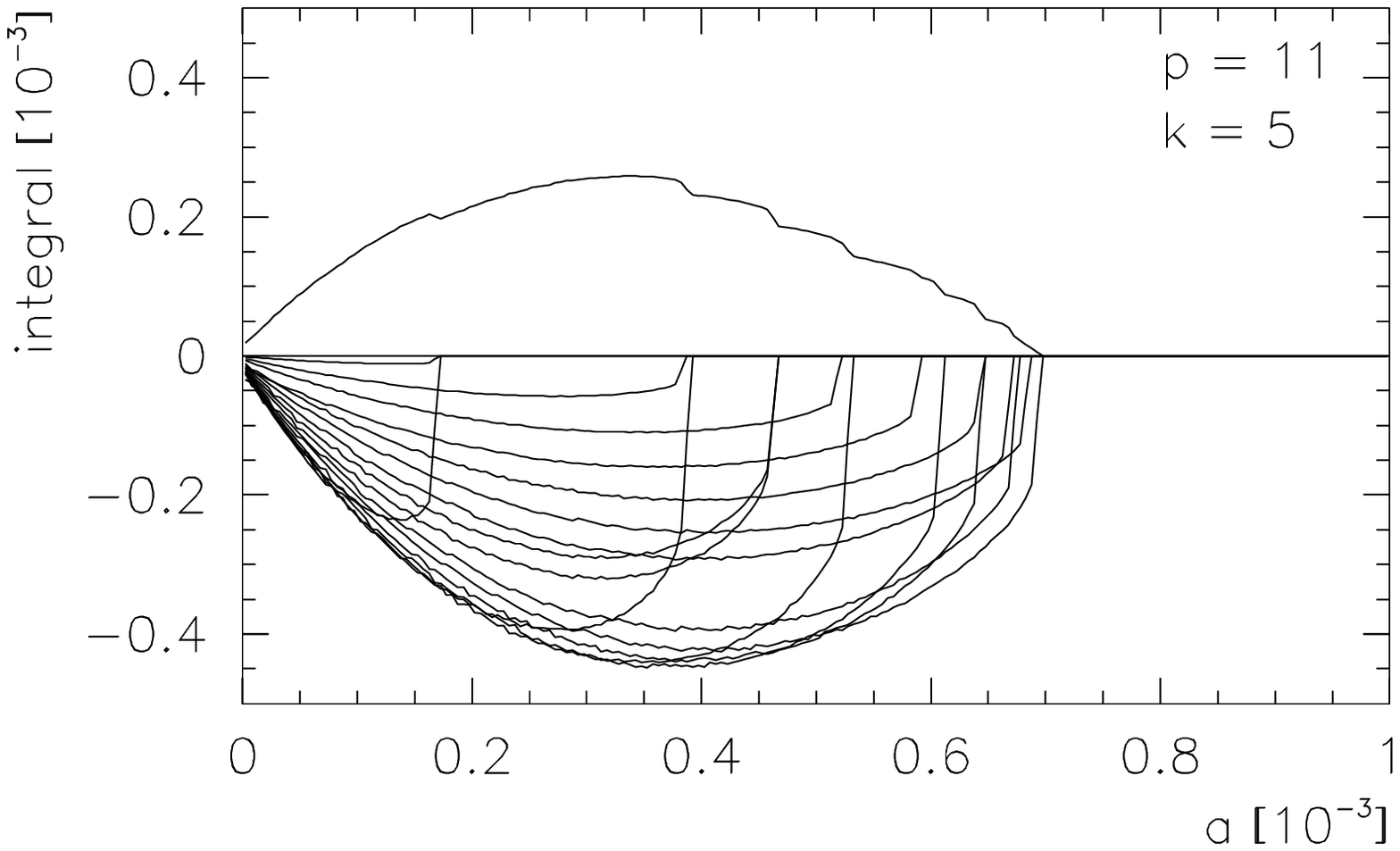, scale=0.8}
\caption{\label{fine2b05b}Contributions and sum of all the partition integrals
for $p=11$ and $k=5$ that vanish in the displayed interval for $a$. For better
visibility, the sum is multiplied by $-2^{1-k}$.}
\end{center}\end{figure}

\section{Renormalisation group approach}
The simultaneous parameter-dependent deformations described by draft and blunt
functions are the basis for our treatment in the following. To understand the
complex $a$-dependence of observables such as the vacuum expectation value, we
need to evaluate the integral~(\ref{expectphi}). This integral, as mentioned
before, is split into many partition integrals. These partition integrals are
defined as the part of the integral~(\ref{expectphi}) where the integration
range is restricted to the interval given by the positions $t_a$ and $t_b$
of two adjacent minima surrounding a decaying maximum with starting position
$t_0$,
\begin{equation}
\langle\phi\rangle_a(t_0)=\frac\pi2\int_{t_a}^{t_b}f(t)\cos\pfrac{\pi t}2dt
\end{equation}
where the integrand is given by
\begin{equation}\label{integrand}
f(t):=\int\rho_0(\phi_+)d\phi_+\rho_0(\phi_-)d\phi_-T_{2a}\Big(\ldots
  T_{2a}\left(\sin(\pi t/2);\phi_++\phi_-\right)\ldots;
  T_{2^p}(\phi_+)+T_{2^p}(\phi_-)\Big).
\end{equation}
In Figs.~\ref{fine2b04a} and~\ref{fine2b05b}, we show a sequence of partition
integrals and their sum for values of $p$ and $k$ with constant difference
$p-k=6$. At first sight one might conjecture that all these partition
integrals vanish at the same maximal value of $a$, called the {\it maximal
reach\/} for a given $(p,k)$. However, our careful analysis shows that due to
different values of the draft function, many partition integrals vanish at
values below the maximal value for a given $(p,k)$.

Even though the draft function does not depend on $p$ and $k$, the effect of
partition integrals to vanish below the maximal reach increases with the order
because the maxima are closer together and are smeared out faster. For this
reason, also the maximal reach of the partition integrals (i.e.\ the value $a$
at which all partition integrals of the same class vanish) is decreasing in
the sequence of the two diagrams shown in Figs.~\ref{fine2b04a}
and~\ref{fine2b05b}.

\begin{figure}\begin{center}
\epsfig{figure=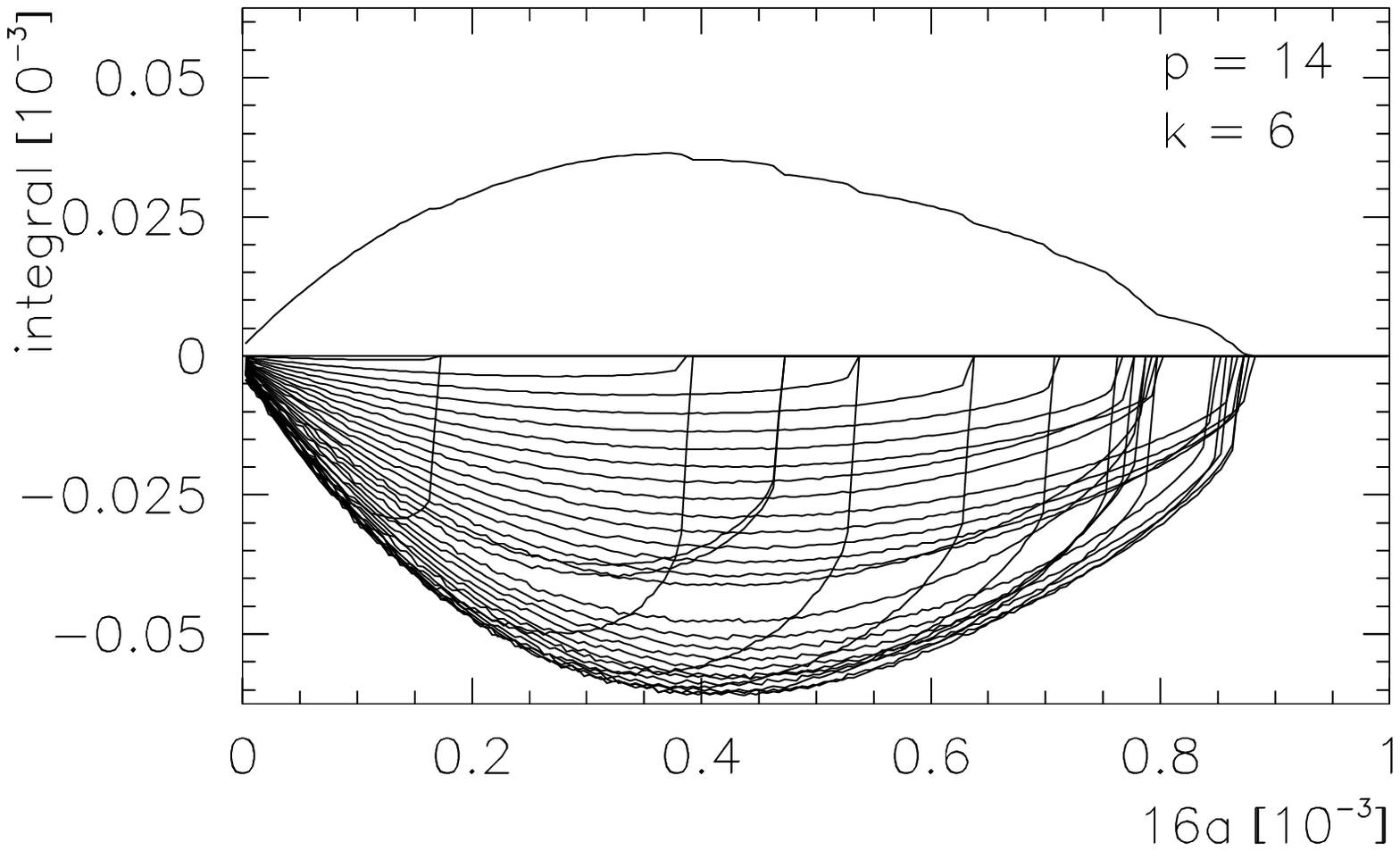, scale=0.8}
\caption{\label{fine2b06e}Contributions and sum of all the partition integrals
for $p=14$ and $k=6$ that vanish in the displayed interval. For better
visibility, the sum is multiplied by $-2^{1-k}$.}
\vspace{36pt}
\epsfig{figure=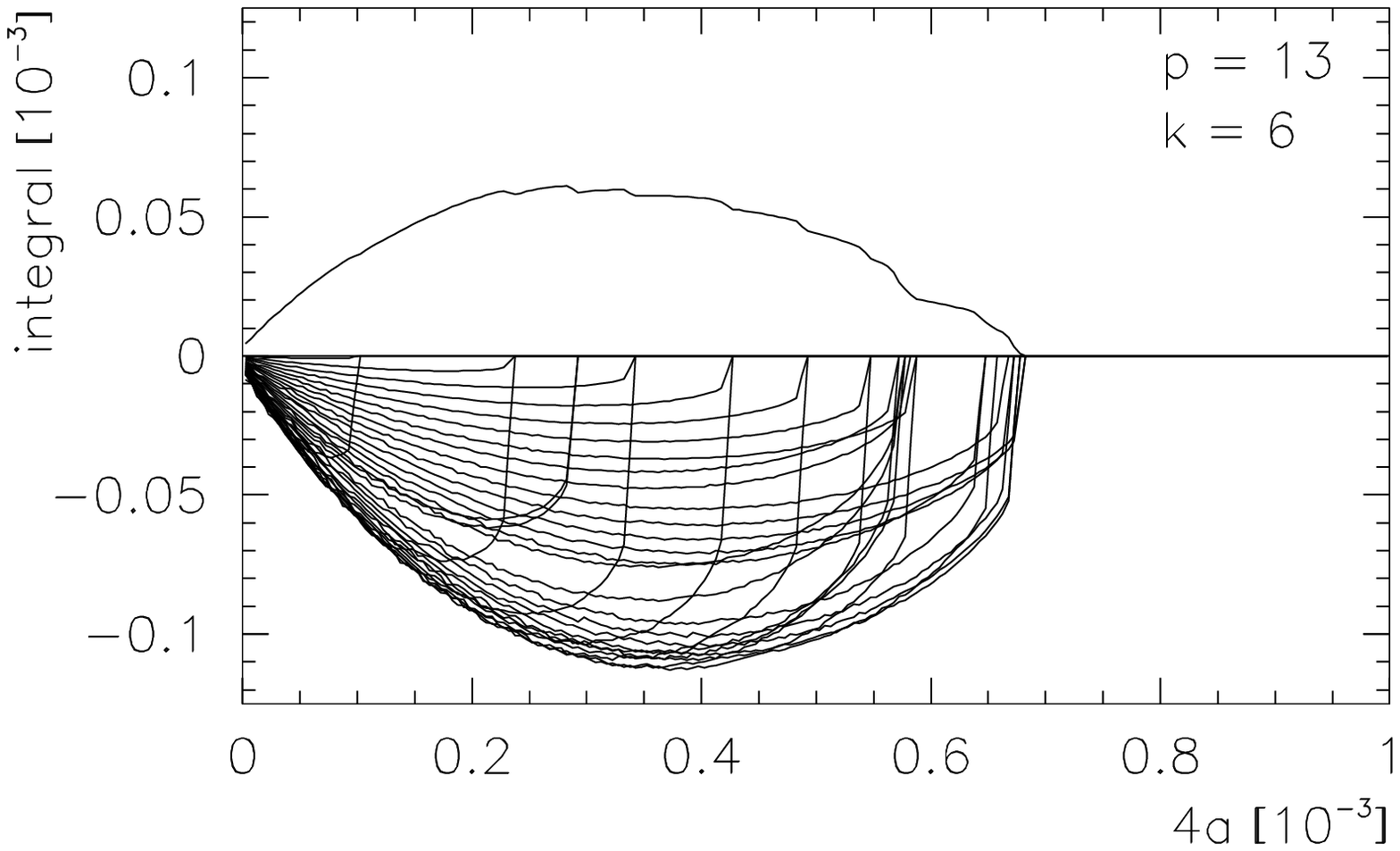, scale=0.8}
\caption{\label{fine2b06d} Same as Fig.~12 but for $p=13$ and $k=6$.}
\end{center}\end{figure}

\vspace{7pt}
In the spirit of a suitable renormalization group theory, we are now looking
for a transition $p\to\infty$ that preserves the shape and sum of the partition
integrals. Comparing the case $p=14$, $k=6$ as displayed in
Fig.~\ref{fine2b06e} with Fig.~\ref{fine2b04a} ($p=10$, $k=4$), and the case
$p=13$, $k=6$ in Fig.~\ref{fine2b06d} with Fig.~\ref{fine2b05b} ($p=11$,
$k=5$), we find that the quantity to be kept constant is given by the
{\em reduced order $\hat p=p-2k$\/} which in the two examples is given by
$\hat p=2$ and $\hat p=1$, respectively. However, in Figs.~\ref{fine2b06e}
and~\ref{fine2b06d} we have stretched the axis for $a$ by a factor of $4$ and
$16$, respectively. Because we are looking at the behaviour in the scaling
region only, this means that instead of the coupling $a$ one should use the
{\em reduced coupling $\hat a=4^{p-k}a$} which for the intervals shown in
Figs.~\ref{fine2b04a}--\ref{fine2b06e} is found in the interval
$\hat a\in[0,4]$. With the above rescaling prescriptions, the overall problem
is formulated in a {\em scale invariant} way in the limit $p\to\infty$.

\subsection{The model function}
The Feigenbaum fixed point function in Feigenbaum's renormalization group
approach for 1-dimensional critical systems can be well approximated close to
the vicinity of the maximum of the map~\cite{feigenbaum1,feigenbaum2,%
beck-schloegl}. Similarly, we can also find a good approximation of the
integrand in our problem if we are close to a decaying maximum. We call this
approximating function of the integrand the `model function'. Using the
analytical expressions for draft and blunt functions, the integrand in
Eq.~(\ref{integrand}) in the region close to a decaying maximum is
approximated by the model function
\begin{eqnarray}\label{model0}
\lefteqn{f(\Delta t;t_0)\ = \int_{-1}^1\rho_0(\phi_+)d\phi_+\rho_0(\phi_-)
  d\phi_-\times\strut}\nonumber\\&&\strut\times
  \cos\left(2^p\sqrt{\pi^2\left(\Delta t-a\tilde d^{(k)}(t_0;\phi_+,\phi_-)
  \right)^2+4^{1-k}ar_2^\infty(-T_{2^k}(\phi_+),-T_{2^k}(\phi_-))}\right)
\end{eqnarray}
which is derived from Eq.~(\ref{blunt}). This turns out to be a quite precise
approximation for $f(t_0+\Delta t)$ in Eq.~(\ref{integrand}) where $\Delta t$
is the deviation from the starting position $t_0$ of the specified maximum. If
in addition to the reduced order and coupling one uses the {\em reduced
deviation $\Delta\hat t=2^p\Delta t\in[-1,1]$}, the limit $p\to\infty$ can be
performed for the model function~(\ref{model0}), rewritten as
\begin{eqnarray}\label{model1}
\lefteqn{\hat f^{(\hat p)}(\Delta\hat t;t_0):=f(2^{-p}\Delta\hat t;t_0)
  \ =\ \int\rho_0(\phi_+)d\rho_+\rho_0(\phi_-)d\phi_-
  \times\strut}\nonumber\\&&\strut\times
  \cos\left(\sqrt{\pi^2\left(\Delta\hat t
  -2^{\hat p}\hat a\tilde d^{(k)}(t_0;\phi_+,\phi_-)\right)^2
  +2\hat ar_2^\infty\left(-T_{2^k}(\phi_+),-T_{2^k}(\phi_-)\right)}\right).
  \qquad
\end{eqnarray}
The function $r_2^\infty(\phi_+,\phi_-)$ uses the $k$-fold Chebyshev map for the
arguments $\phi_\pm$ while the arguments of the draft function are just
$\phi_\pm$. Keeping the reduced order $\hat p=p-2k$ constant, the class $k$
increases together with the order $p$. Because of this, the high-frequency
blunt can be separated from the low-frequency draft, leading to an effective
blunt of $8/3$ (see Appendix~C). The final result for the model function in
the limit $p\to\infty$ reads
\begin{equation}\label{model2}
\hat f^{(\hat p)}(\Delta\hat t;t_0)=\int\rho_0(\phi_+)d\phi_+
  \rho_0(\phi_-)d\phi_-\cos\left(\sqrt{\pi^2\left(\Delta\hat t
  -2^{-\hat p}\hat a\tilde d^{(k)}(t_0;\phi_+,\phi_-)\right)^2
  +\frac{8\hat a}3}\right).
\end{equation}

\subsection{Fourier transformation}
We note that the integrand in the above equation~(\ref{model2}) is essentially
of the form $g(a,t):=\cos(\sqrt{t^2+a})$. Inspired by quantum field theory, we
proceed to a Fourier transform denoted as
\begin{equation}
\tilde g(a,\omega)=\int_{-\infty}^\infty\cos(\sqrt{t^2+a})\cos(\omega t)dt.
\end{equation}
It can be shown that $g(a,\omega)=0$ for $\omega^2>1$, but otherwise
\begin{equation}
\tilde g(a,\omega)=2\pi\delta(1-\omega^2)-\frac{\pi a}{\sqrt{a(1-\omega^2)}}
  J_1\left(\sqrt{a(1-\omega^2)}\right),
\end{equation}
where $J_1(z)$ is the Bessel function of the first kind. One obtains
\begin{equation}
g(a,t)=\cos t-\int_{-1}^{+1}\frac{a\,d\omega}{2\sqrt{a(1-\omega^2)}}
  J_1\left(\sqrt{a(1-\omega^2)}\right)\cos(\omega t).
\end{equation}
If this is inserted into Eq.~(\ref{model2}), the model function can be
expressed as
\begin{eqnarray}\label{model3}
\lefteqn{f^{(\hat p)}(\Delta\hat t;t_0)\ =\ \int\rho_0(\phi_+)d\phi_+
  \rho_0(\phi_-)d\phi_-\Bigg[\cos\left(i\pi\left(\Delta\hat t
  -2^{-\hat p}\hat a\tilde d^{(k)}(t_0;\phi_+,\phi_-)\right)\right)}
  \nonumber\\&&\strut
  -\int_{-1}^{+1}\frac{4\hat a\,d\omega}{3\sqrt{8\hat a(1-\omega^2)/3}}
  J_1\left(\sqrt{8\hat a(1-\omega^2)/3}\right)\cos\left(i\pi\omega
  \left(\Delta\hat t-2^{-\hat p}\hat a\tilde d^{(k)}(t_0;\phi_+,\phi_-)
  \right)\right)\Bigg].\qquad
\end{eqnarray}

\subsection{Bessel integrals and Feynman webs}
\begin{figure}\begin{center}
\epsfig{figure=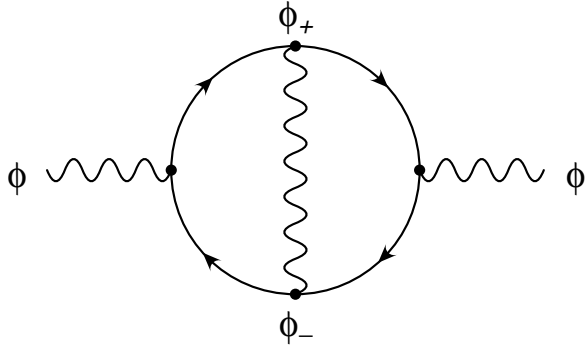, scale=0.5}
\caption{\label{fishdiag}A so-called fish-type diagram for the correlator
between bosons (twiggles) and fermions (straight lines)}
\end{center}\end{figure}

The occurence of the Bessel function of degree one discloses a surprising
connection to the Feynman diagrams of elementary particle physics. In
calculating Feynman diagrams in configuration space and afterwards returning
to momentum space with Euclidean metrics in $D$ dimensions, the exponential
function can be expanded into Gegenbauer polynomials
$C_l^\lambda(w)$~\cite{Groote:2005ay},
\begin{equation}
e^{ip\cdot x}=\Gamma(\lambda)\pfrac{px}2^{-\lambda}\sum_{l=0}^\infty
i^l(\lambda+l)J_{\lambda+l}(px)C_l^\lambda\pfrac{p\cdot x}{px},
\end{equation}
where $D=2\lambda+2$. Note that $p\cdot x=p_\mu x^\mu$ is the scalar product
of two $D$-dimensional vectors, while $px$ is the product of the two absolute
values $p=\sqrt{p_\mu p^\mu}$ and $x=\sqrt{x_\mu x^\mu}$. If the integrand
does not depend on the $D$-dimensional vectors but only on their absolute
values, then the $D$-dimensional integration measure of the Fourier
transformation simplifies to
\begin{equation}
\int e^{ip\cdot x}d^Dx=2\pi^{\lambda+1}\int_0^\infty
  \pfrac{px}2^{-\lambda}J_\lambda(px)x^{2\lambda+1}dx.
\end{equation}
An interesting observation is that the model function contains the Bessel
function (and a similar factor in front of it) of degree $\lambda=1$, which
corresponds to $D=2+2=4$ space-time dimensions. If one aims for a physical
embedding of the chaotic string dynamics into generalized versions of quantum
field theory~\cite{Beck:2002}, then this means that $\lambda=1$ {\em cites
four-dimensional space-time without the explicit emergence of this space-time
itself\/}. If we interpret the integrand of the Bessel function as a
correlator function in (obviously, four-dimensional) configuration space, the
integration over $\phi_+$ and $\phi_-$ basically means the insertion of two
vertices into the diagram which are connected by a propagator, i.e.\ a virtual
particle transition. The fish-type diagram in Fig.~\ref{fishdiag}, however,
is a compactified version of the Feynman web that was used in
Ref.~\cite{Beck:2002} to give a physical interpretation to the chaotic string
dynamics.

\section{Limiting cases}
Clearly the exact renormalization group treatment requires the limit
$p\to\infty$. On the other hand, of relevance for a scale invariant
formulation is the reduced order $\hat{p}=p-2k$. In the following we look at
different limiting behavior depending on the relative size of $p$ and $k$. It
turns out that one can distinguish two different limit ranges which we call
the perturbative and the nonperturbative range.

\begin{figure}\begin{center}
\epsfig{figure=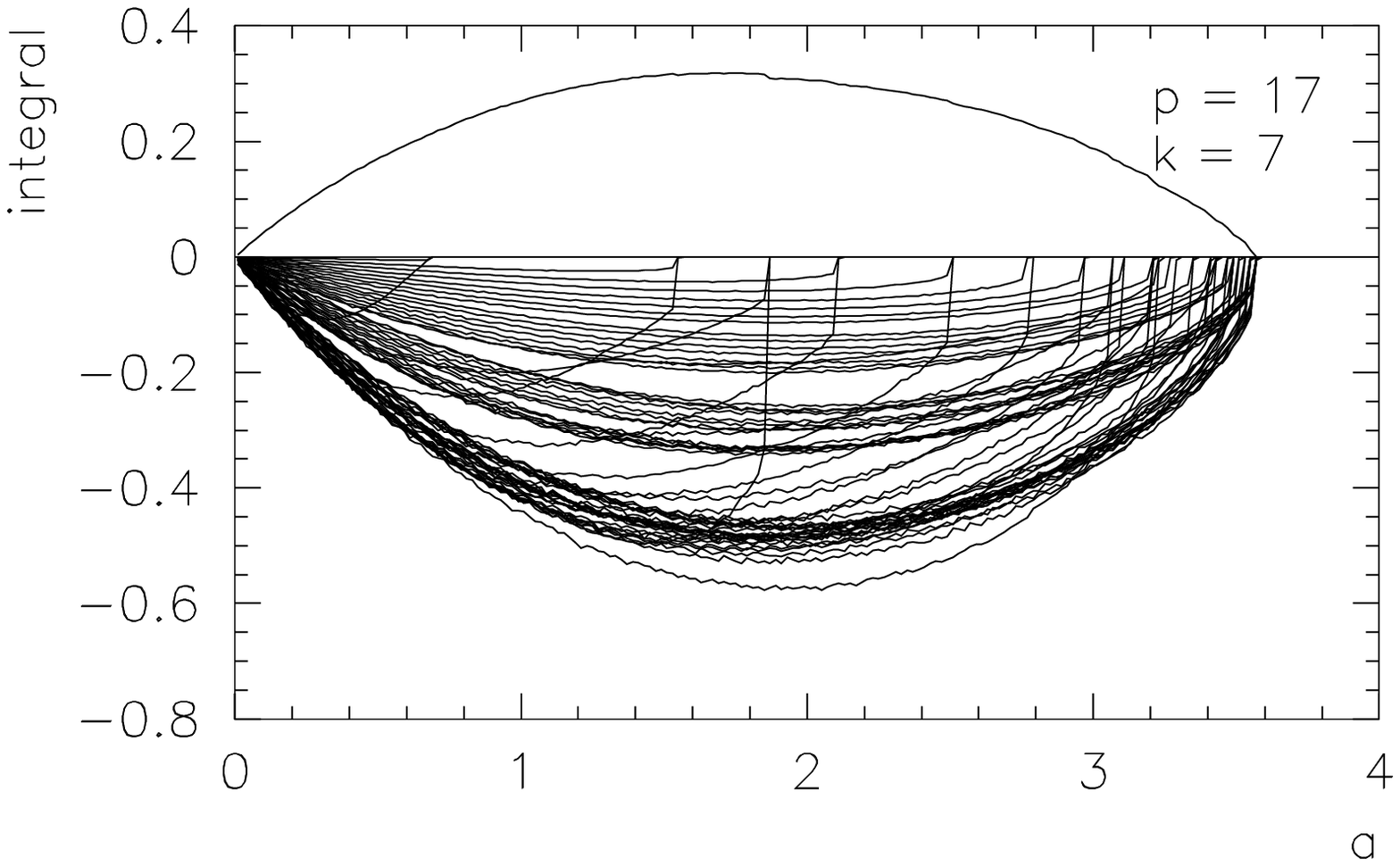, scale=0.6}\\[12pt]
\epsfig{figure=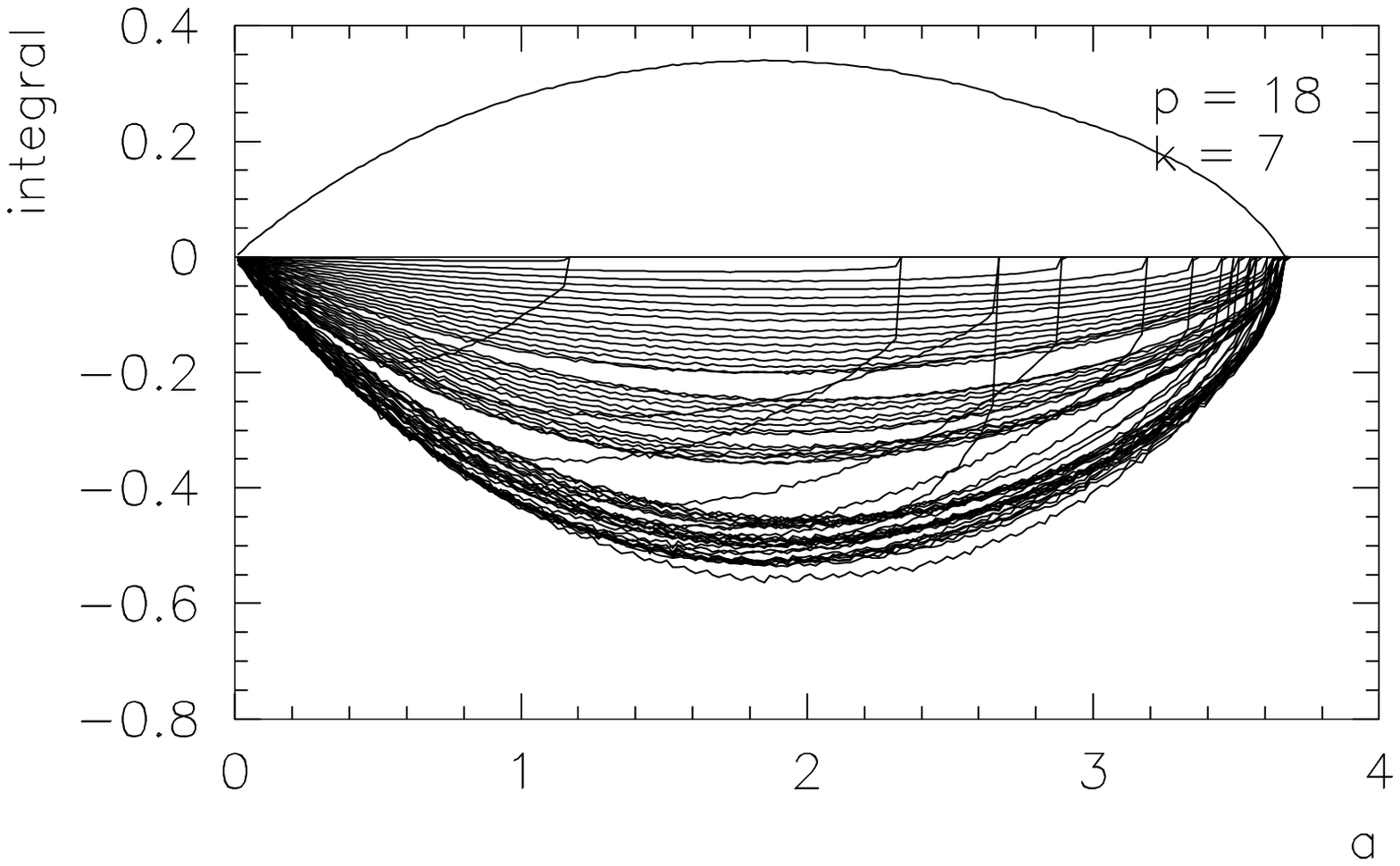, scale=0.6}\\[12pt]
\epsfig{figure=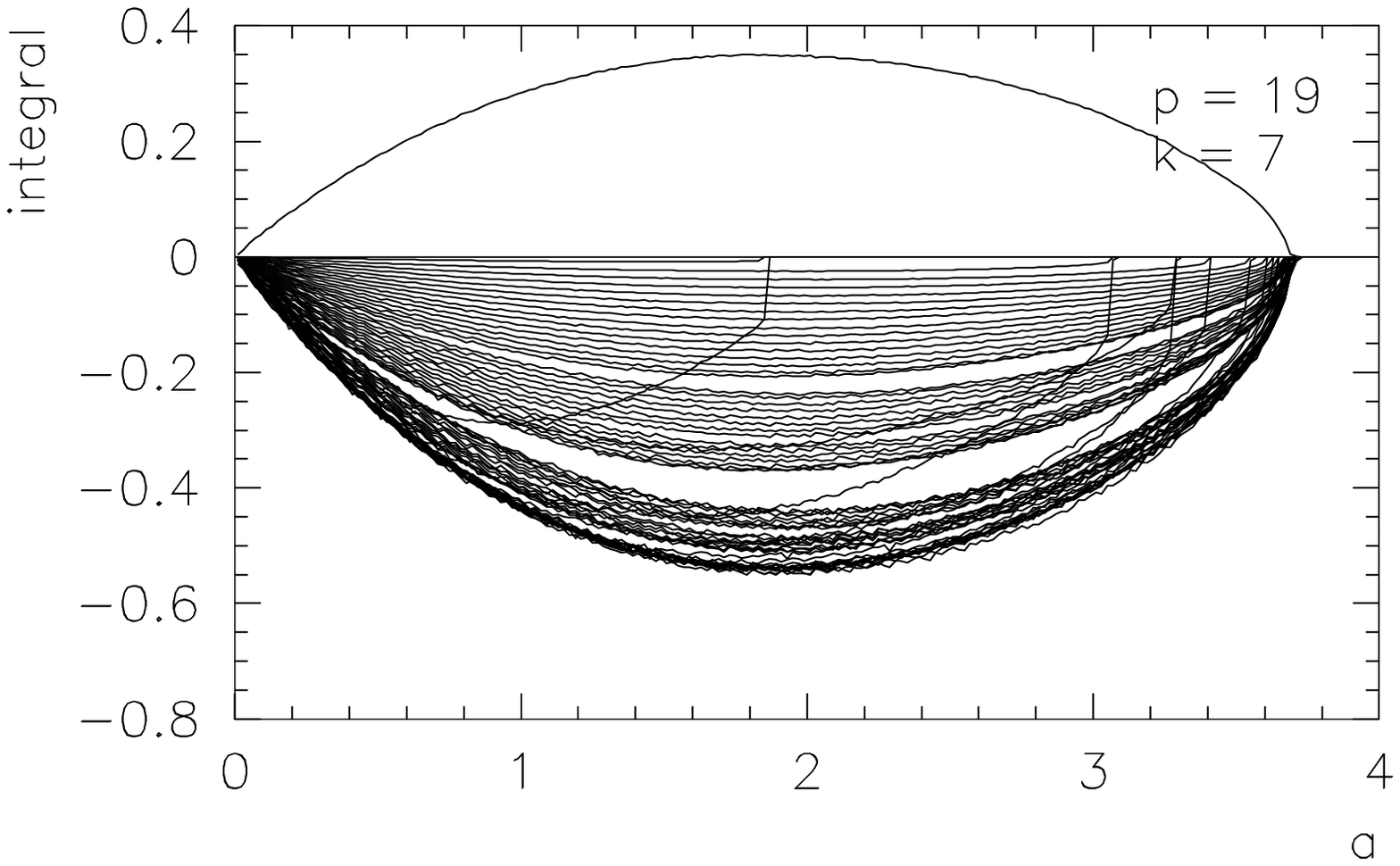, scale=0.6}
\caption{\label{fine2b0x7}Contributions and sum of partition integrals of
class $k=7$ for increasing order $p=17,18,19$ (from top to bottom,
corresponding to the reduced order $\hat p=3,4,5$) as functions of $\hat a$.
For better visibility, the sum is multiplied by $-2^{1-k}$.}
\end{center}\end{figure}

\subsection{The perturbative range}
Increasing the reduced order $\hat p$, the maximal reach does not increase in
the same way. Instead, more and more partition integrals vanish close to the
maximal reach (see Fig~\ref{fine2b0x7}). This means that compared to the blunt
effect, the draft effect can be neglected. Note the similarity of the three
displayed pictures in Fig.~\ref{fine2b0x7}, which is due to the fact that the
reduced coupling $\hat{a}=4^{p-k}a$ is used. This range of parameters $p$ and
$k$ does not contribute to the fine structure but is responsible for the
global selfsimilar behaviour of expectations of observables.

\begin{figure}\begin{center}
\epsfig{figure=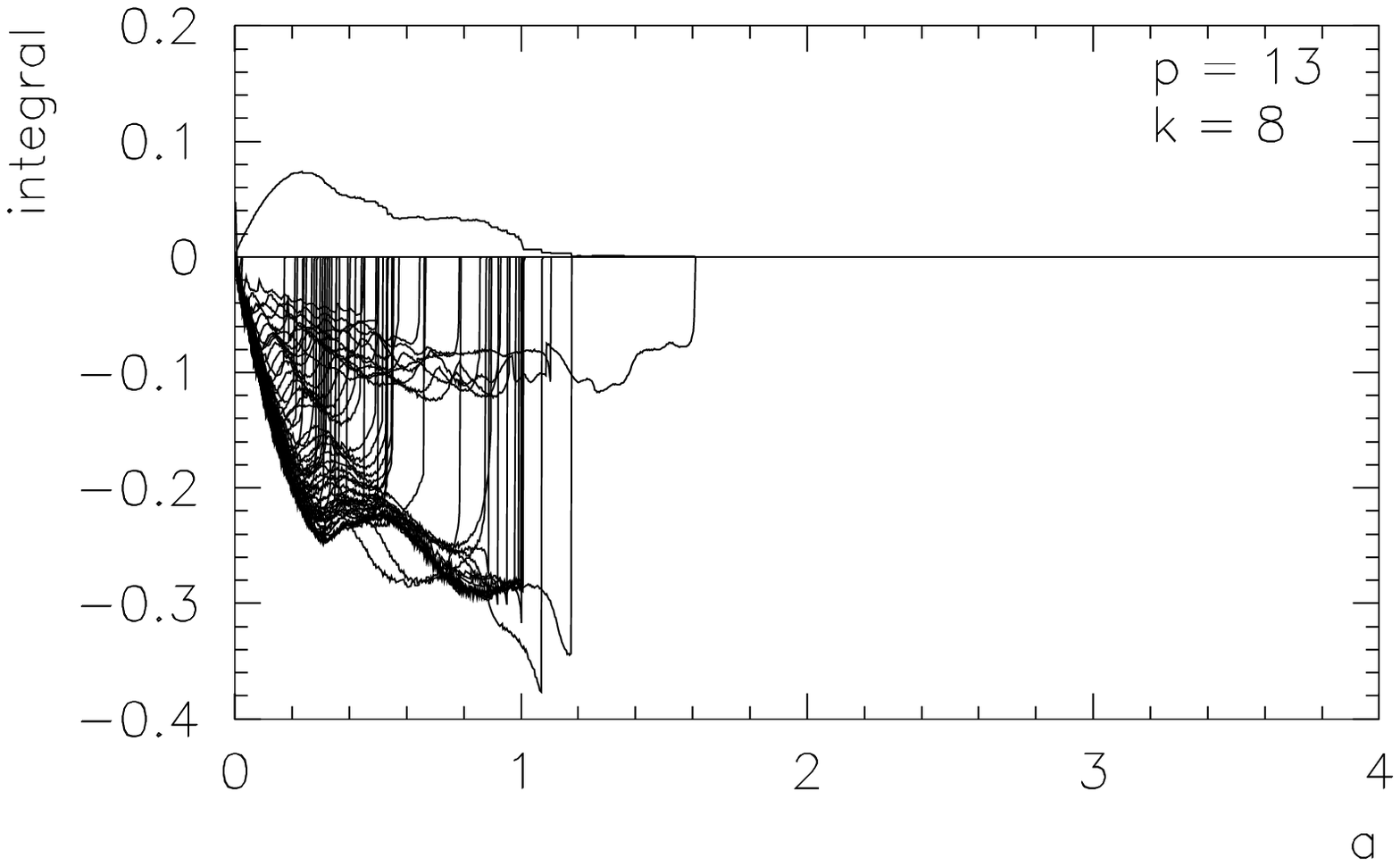, scale=0.8}
\caption{\label{fine2b08d}Contributions and sum of partition integrals
for $p=13$ and $k=8$ as a function of $\hat a\in[0,4]$. For better visibility,
the sum is multiplied by $-2^{1-k}$.}
\end{center}\end{figure}

\subsection{The nonperturbative range}
More interesting for the fine structure is the case where the reduced order
$\hat p$ becomes negative. Fig.~\ref{fine2b08d} unveils a rich structure.
While Fig.~\ref{fine2b08d} is calculated by using the model
function~(\ref{model3}), the calculation for the open string of length $3$
leads to the same result. Finally, the result for class $k=9$ (with the same
reduced order $\hat p=-3$, or $p=15$) is again very similar. This means that
the model function correctly describes the effects seen in this
nonperturbative range. Very characteristic is the kink close to $\hat a=0.5$
which is relevant for the fine structure. This kink has its origin in a
majority of partition integrals swinging twice before they vanish.

\subsection{Development of bounded states}
Characteristic for the nonperturbative range is that for a given decaying
maximum, the neighbouring maxima are influenced strongly by the decay of the
central maximum. They decay nearly as fast as the central maximum. If we use
as an analogy the language of particle physics, then maxima seem to interact
and to formate something like a bounded state. An interesting open question is
whether these ``bounded states'' can be described properly within our current
formalism based on partitions, or whether a different approach is necessary
for this.

\section{The exponential draft function}
In the following we will perform further calculations on the draft function,
which will ultimately lead to the implementation of graph theoretical
methods previously introduced in Refs.~\cite{Beck:1991,Hilgers:2001}.

Our motivation is as follows. Higher-order correlations of uncoupled Chebychev
maps have been previously calculated using a graph-theoretical method. The
graphs relevant for $N$-th order Chebychev maps consist of forests of $N$-ary
trees~\cite{Beck:1991,Hilgers:2001}. It is now intriguing that the same type
of graphs can be used for the coupled map system when calculating the relevant
scaling functions. This reminds us of the use of Feynman graphs in quantum
field theory, where one also uses graphs up to a certain order to understand
the coupled system. 

Let us define an {\em exponential draft function} by
\begin{equation}\label{expdraft}
\tilde h(\omega;t_0):=\int_{-1}^{+1}\exp\left(-i\omega\tilde d^{(k)}
(t_0;\phi_+)\right)\rho_0(\phi_+)d\phi_+.
\end{equation}
With the help of this the model function can be written as
\begin{eqnarray}\label{model4}
\lefteqn{f^{(\hat p)}(\Delta t;t_0)\ =\ \real\left(e^{i\pi\Delta\hat t}
  \tilde h^2(2^{-\hat p-1}\pi\hat a;t_0)\right)}\nonumber\\&&\strut
  -\int_{-1}^{+1}\frac{4\hat a\,d\omega}{3\sqrt{8\hat a(1-\omega^2)/3}}
  J_1\left(\sqrt{8\hat a(1-\omega^2)/3}\right)\real\left(e^{i\pi\omega\Delta
  \hat t}\tilde h^2(2^{-\hat p-1}\pi\hat a\omega;t_0)\right).\qquad
\end{eqnarray}
The relative minima (and maxima) of the model function which define the
boundaries of the partition integrals are found by calculating the first
derivative with respect to $\Delta\hat t$ and determining the zeros of this
derivative. One obtains
\begin{eqnarray}\label{model4d}
\lefteqn{\imag\left(e^{i\pi\Delta\hat t}
  \tilde h^2(2^{-\hat p-1}\pi\hat a;t_0)\right)\ =}\nonumber\\
    &=&\int_0^1\frac{8\hat a\omega\,d\omega}{3\sqrt{8\hat a(1-\omega^2)/3}}
  J_1\left(\sqrt{8\hat a(1-\omega^2)/3}\right)\imag\left(e^{i\pi\omega\Delta
  \hat t}\tilde h^2(2^{-\hat p-1}\pi\hat a\omega;t_0)\right),\qquad
\end{eqnarray}
using the fact that the integrand is an even function of $\omega$.

\begin{figure}\begin{center}
\epsfig{figure=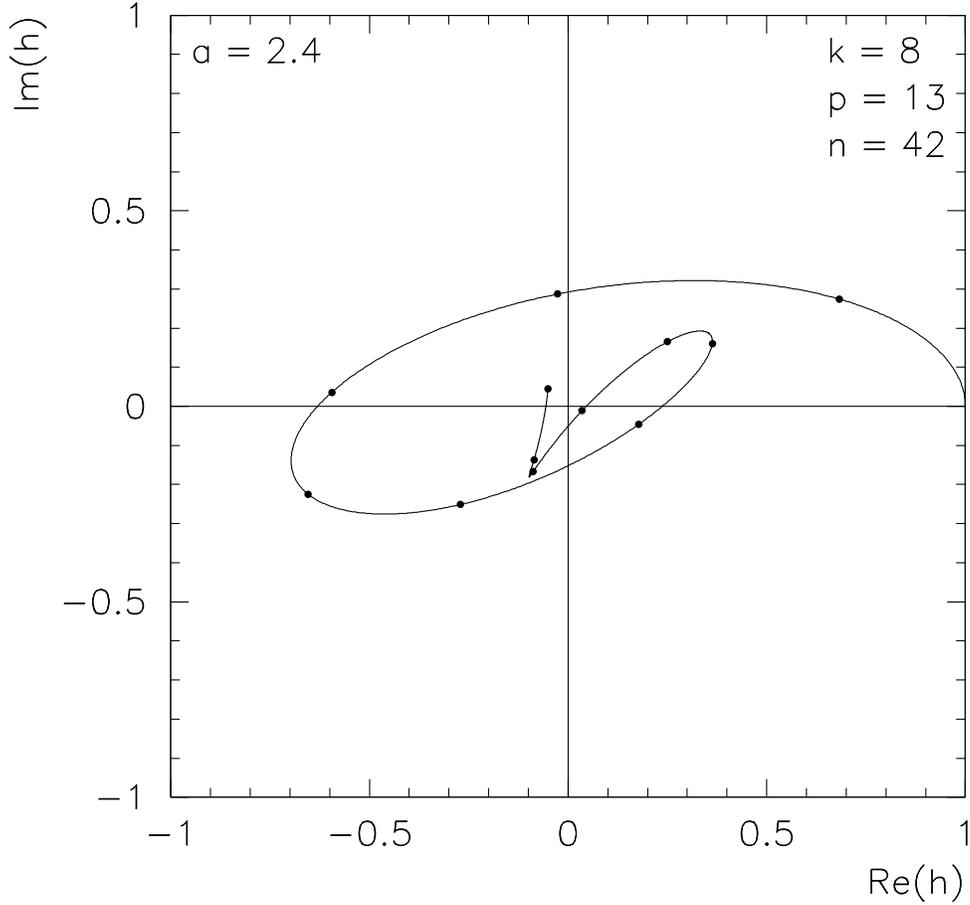, scale=0.8}
\caption{\label{omega2b08d2a}Parametric curve of the exponential draft
function $\tilde h(2^{-\hat p-1}\pi\hat a\omega;t_0)$ for $k=8$, $p=13$ and
$n=42$ with reduced coupling $\hat a=2.4$ and $\omega\in[0,1]$. The twelve
points represent 12 equidistant values of $\omega$ for a fixed value of
$\hat a=2.4$.}
\end{center}\end{figure}

\vspace{7pt}
The exponential draft function can be visualised by a parametric curve in the
complex plane, as it is shown for $k=8$, $p=13$ and a particular partition
integral specified by $n=42$ in Fig.~\ref{omega2b08d2a}. The long range of the
curve is due to the fact that $2^{-\hat p-1}\pi\hat a=4\pi\hat a$ is not a
small number. Note that the curve for $\omega\in[-1,0]$ is obtained by
mirroring at the real axis. The parametric curve shows quite an irregular
behaviour which is hard to predict.

\subsection{Analytical expression for the exponential draft function}
One can perform a concrete calculation to obtain the exponential draft
function using graph theory. Inserting Eq.~(\ref{dk1}) into
Eq.~(\ref{expdraft}), one obtains\footnote{Without loss of generality,
$T_{2^l}(-\phi_\pm)$ can be replaced by $T_{2^l}(\phi_\pm)$ under the
integration.}
\begin{eqnarray}\label{edraft}
\tilde h(\omega;t_0)&=&e^{-i\omega\tilde d^{(k)}(t_0)}
  \int_{-1}^1\exp\left(-i\omega\sum_{l=0}^{k-1}T_{2^l}(\phi_+)
  \tilde d^{(k)}_l(t_0)\right)\rho(\phi_+)d\phi_+\ =\nonumber\\
  &=&e^{-i\omega\tilde d^{(k)}(t_0)}\sum_{n=0}^\infty\frac{(-i\omega)^n}{n!}
  \int_{-1}^1\Big(\sum_{l=0}^{k-1}T_{2^l}(\phi_+)\tilde d^{(k)}_l(t_0)\Big)^n
  \rho_0(\phi_+)d\phi_+.
\end{eqnarray}
In the following we use the short hand notations
$d:=\tilde d^{(k)}(t_0)$ and $d_l:=\tilde d^{(k)}_l(t_0)$ and the convention that
sums run over values where all indices under the sum take nonnegative values
less than $k$. By performing the integration over $\phi_+$ one obtains
\begin{eqnarray}\label{expseries}
\lefteqn{\tilde h(\omega;t_0)\ =\ e^{-i\omega d}
  \Bigg(1+\frac{(-i\omega)^2}4\sum d_l^2
  +\frac{(-i\omega)^3}8\sum d_l^2d_{l+1}}\nonumber\\&&\strut
  +\frac{(-i\omega)^4}{16}\sum d_l^2d_{l+1}d_{l+2}
  +\frac{(-i\omega)^4}{32}\Bigg(\sum d_l^2\Bigg)^2
  -\frac{(-i\omega)^4}{64}\sum d_l^4\nonumber\\&&\strut
  +\frac{(-i\omega)^5}{32}\sum d_l^2d_{l+1}d_{l+2}d_{l+3}
  +\frac{(-i\omega)^5}{32}\sum d_l^2d_{l+1}\sum d_{l'}^2
  +\frac{(-i\omega)^5}{384}\sum d_l^4d_{l+2}\nonumber\\&&\strut
  -\frac{(-i\omega)^5}{64}\sum d_l^2d_{l+1}^3
  -\frac{(-i\omega)^5}{48}\sum d_l^4d_{l+1}
  +\frac{(-i\omega)^6}{64}\sum d_l^2d_{l+1}d_{l+2}d_{l+3}d_{l+4}
  \nonumber\\&&\strut
  +\frac{(-i\omega)^6}{384}\sum d_l^2d_{l+1}^3d_{l+3}
  +\frac{(-i\omega)^6}{768}\sum d_l^4d_{l+2}d_{l+3}
  -\frac{(-i\omega)^6}{48}\sum d_l^2d_{l+1}^3d_{l+2}
  \nonumber\\&&\strut
  -\frac{(-i\omega)^6}{96}\sum d_l^4d_{l+1}d_{l+2}
  -\frac{(-i\omega)^6}{128}\sum d_l^2d_{l+1}d_{l+2}^3
  -11\frac{(-i\omega)^6}{1536}\sum d_l^4d_{l+1}^2
  \nonumber\\&&\strut
  +\frac{(-i\omega)^6}{64}\sum d_l^2d_{l+1}d_{l+2}
  \sum_{l'=0}^{k-1}d_{l'}^2
  +\frac{(-i\omega)^6}{128}\Bigg(\sum d_l^2d_{l+1}\Bigg)^2
  +\frac{(-i\omega)^6}{576}\sum d_l^6
  \nonumber\\&&\strut
  -\frac{(-i\omega)^6}{256}\sum d_l^4\sum d_{l'}^2
  +\frac{(-i\omega)^6}{384}\Bigg(\sum d_l^2\Bigg)^3+\ldots\ \Bigg).
\end{eqnarray}
Expressing the higher-order correlation of (uncoupled) Chebyshev maps in terms
of a graph theoretical approach~\cite{Beck:1991,Hilgers:2001}, the various
contributions above represent possible integer partitions of powers of $2$ and
can be represented as binary forests. It can be shown that the series can be
resummed as an exponential function
$\tilde h(\omega;t_0)=\exp(i\tilde h_z(\omega;t_0))$ where the exponent
$\tilde h_z(\omega;t_0)$ contains only trees,
\begin{eqnarray}
\lefteqn{\tilde h_z(\omega;t_0)\ =\ -i\omega d
  +\frac{(-i\omega)^2}4\sum d_l^2
  +\frac{(-i\omega)^3}8\sum d_l^2d_{l+1}}\nonumber\\&&\strut
  +\frac{(-i\omega)^4}{16}\sum d_l^2d_{l+1}d_{l+2}
  -\frac{(-i\omega)^4}{64}\sum d_l^4
  +\frac{(-i\omega)^5}{32}\sum d_l^2d_{l+1}d_{l+2}d_{l+3}\nonumber\\&&\strut
  +\frac{(-i\omega)^5}{384}\sum d_l^4d_{l+2}
  -\frac{(-i\omega)^5}{64}\sum d_l^2d_{l+1}^3
  -\frac{(-i\omega)^5}{48}\sum d_l^4d_{l+1}\nonumber\\&&\strut
  +\frac{(-i\omega)^6}{64}\sum d_l^2d_{l+1}d_{l+2}d_{l+3}d_{l+4}
  +\frac{(-i\omega)^6}{384}\sum d_l^2d_{l+1}^3d_{l+3}
  +\frac{(-i\omega)^6}{768}\sum d_l^4d_{l+2}d_{l+3}\nonumber\\&&\strut
  -\frac{(-i\omega)^6}{48}\sum d_l^2d_{l+1}^3d_{l+2}
  -\frac{(-i\omega)^6}{96}\sum d_l^4d_{l+1}d_{l+2}
  -\frac{(-i\omega)^6}{128}\sum d_l^2d_{l+1}d_{l+2}^3\nonumber\\&&\strut
  -11\frac{(-i\omega)^6}{1536}\sum d_l^4d_{l+1}^2
  +\frac{(-i\omega)^6}{576}\sum d_l^6+\ldots
\end{eqnarray}

\subsection{Single trees and multiple trees}
The idea of using binary forest graphs for our calculation can be summarized
as follows: Let $d_l$ be the coefficient of the Chebyshev polynomial
$T_{2^l}(\phi_+)$. Under the replacement $\phi_+=\sin(\pi t_+/2)$ one has
$T_{2^0}(\phi_+)=\sin(\pi t_+/2)$, $T_{2^1}(\phi_+)=-\cos(\pi t_+)$, and
$T_{2^l}(\phi_+)=\cos(2^{l-1}\pi t_+)$ for $l>1$. Therefore, under this
replacement, the Chebyshev poynomials can be represented using complex
exponential functions $\exp(\pm 2^{l-1}i\pi t_+)$. In calculating the integral
in Eq.~(\ref{edraft}), a product of exponential functions corresponding to the
coefficients $d_{l+i}$ in the sums in Eq.~(\ref{expseries}) is integrated over
$\phi_+$. However, the integration of this product gives a non-vanishing
contribution only if the exponent of the complex exponential vanishes
(for details, cf.\ Refs.~\cite{Beck:1991,Hilgers:2001}).

\vspace{7pt}
To give an example, the $l$-th contribution to $\sum d_l^2d_{l+1}$ in
Eq.~(\ref{expseries}) is given by integrals of the exponential function
\begin{equation}
\exp\left((\pm 2^l\pm 2^l\pm 2^{l+1})i\pi t_+\right)
  =\exp\left(2^l(\pm 2^0\pm 2^0\pm 2^1)i\pi t_+\right).
\end{equation}
This integral gives a non-vanishing contribution only in the case where
$\pm 2^0\pm 2^0\pm 2^1=0$ (all the signs being independent). The only
possibilities in this case are $2^0+2^0-2^1=1+1-2=0$ and
$-2^0-2^0+2^1=-1-1+2=0$ which are different only by a global sign. One can
write $2^0+2^0=1+1=2=2^1$ which is a unique partition of the highest power.
In general, the {\em binary sum condition\/} is given by
\begin{equation}
\sum_{m=0}^i\sum_{n=1}^{n_m}s_n2^m=0
\end{equation}
where $s_n=\pm 1$ are the possible signs in the binary sum condition. The same
powers of $2$ in this condition are represented by points (leaves) in the same
layer of a binary forest of graphs. In order to understand these graphs and
their construction methods, one starts with the highest power. The highest
power $2^i$ can be decomposed into lower powers by using $2^i=2^{i-1}+2^{i-1}$.
This is represented by two lines going down from this (highest) point of the
graph. If both lower powers are found in the binary sum condition (like in
$2^1=2^0+2^0$ in the example above), the lines are closed by points (leaves are drawn) and a tree is created. If one (or both) of the lower powers are not
found in the vanishing sum, this lower power again is decomposed in smaller
powers, until the powers are met by contributions in the binary sum condition.

\vspace{7pt}
The main part of the exponent $\tilde h_z(\omega;t_0)$ contains single trees,
corresponding to unique decompositions of the highest power of $2$. They are
given by
\begin{eqnarray}
n=2:&&\sum d_l^2\quad\mbox{means $1-1=0$}\qquad
  \epsfig{figure=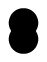, scale=0.5}\nonumber\\
n=3:&&\sum d_l^2d_{l+1}\quad\mbox{means $1+1-2=0$}\qquad
  \epsfig{figure=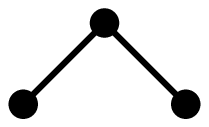, scale=0.5}\nonumber\\
n=4:&&\sum d_l^2d_{l+1}d_{l+2}\quad\mbox{means $1+1+2-4=0$}\qquad
  \epsfig{figure=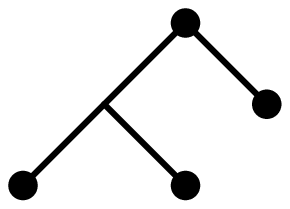, scale=0.5}
  \nonumber\\
n=5:&&\sum d_l^2d_{l+1}d_{l+2}d_{l+3}\quad
  \mbox{means $1+1+2+4-8=0$}\qquad
  \epsfig{figure=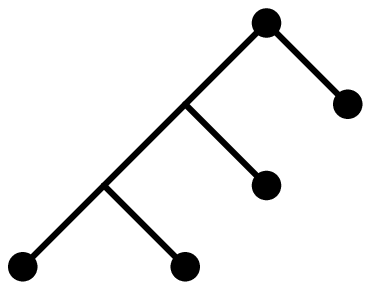, scale=0.5}\nonumber\\
    &&\sum d_l^4d_{l+2}\quad
  \mbox{means $1+1+1+1-4=0$}\qquad
  \epsfig{figure=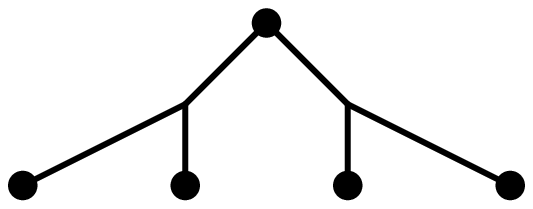, scale=0.5}\nonumber\\
n=6:&&\sum d_l^2d_{l+1}d_{l+2}d_{l+3}d_{l+4}\quad
  \mbox{means $1+1+2+4+8-16=0$}\qquad
  \epsfig{figure=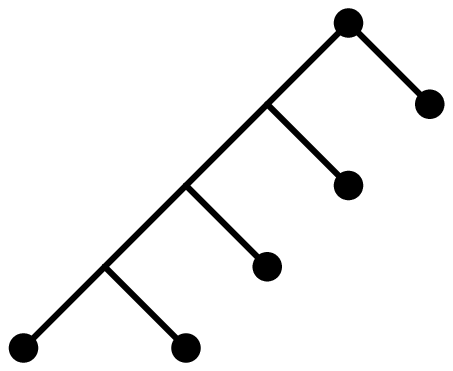, scale=0.5}\nonumber\\
    &&\sum d_l^4d_{l+2}d_{l+3}\quad
  \mbox{means $1+1+1+1+4-8=0$}\qquad
  \epsfig{figure=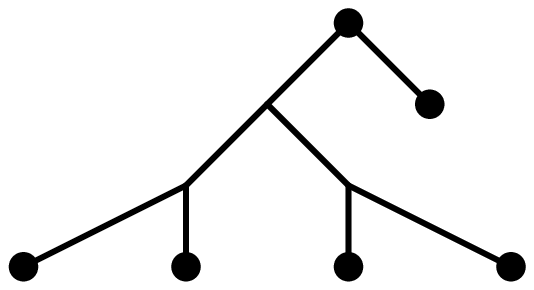, scale=0.5}\nonumber\\
    &&\sum d_l^2d_{l+1}^3d_{l+3}\quad
  \mbox{means $1+1+2+2+2-8=0$}\qquad
  \epsfig{figure=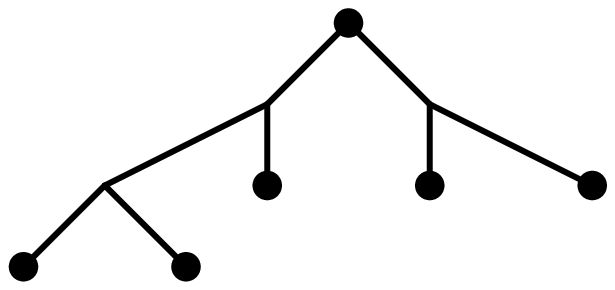, scale=0.5}\qquad
\end{eqnarray}
The coefficients for these contributions are given by
\begin{equation}
\frac1{2^{n-1}n_0!n_1!\cdots n_{n-2}!}
\end{equation}
where $n_k$ counts the occurence of $2^k$ in the decomposition. However, there
are also multiple trees, like for instance $\sum d_l^2d_{l+1}^3$, corresponding
to $+1+1-2|+2-2=0$ where the vertical bar separates the two different
independent relations $+1+1-2=0$ and $+2-2=0$. In order to deal with these
contributions and their coefficients, it is appropriate to work with so-called
summation ordering (for more details see Appendix~D).

\subsection{Summation ordering}
If for multiple sums one excludes repeating indices, one ends up with a
summation ordered expression, similar to time ordered expressions in quantum
field theory. One obtains
\begin{eqnarray}
\lefteqn{\tilde h(\omega;t_0)\ =\ e^{-i\omega d}\bigg(1
  +\frac{(-i\omega)^2}4\sum d_l^2
  +\frac{(-i\omega)^3}8\sum d_l^2d_{l+1}
  +\frac{(-i\omega)^4}{16}\sum d_l^3d_{l+1}d_{l+2}}\nonumber\\&&\strut
  +\frac{(-i\omega)^4}{32}\sum d_l^2\sum'd_{l'}^2
  +\frac{(-i\omega)^4}{64}\sum d_l^4
  +\frac{(-i\omega)^5}{32}\sum d_l^2d_{l+1}d_{l+2}d_{l+3}\nonumber\\&&\strut
  +\frac{(-i\omega)^5}{32}\sum d_l^2d_{l+1}\sum'd_{l'}^2
  +\frac{(-i\omega)^5}{384}\sum d_l^4d_{l+2}
  +\frac{(-i\omega)^5}{64}\sum d_l^2d_{l+1}^3\nonumber\\&&\strut
  +\frac{(-i\omega)^5}{96}\sum d_l^4d_{l+1}
  +\frac{(-i\omega)^6}{64}\sum d_l^2d_{l+1}d_{l+2}d_{l+3}d_{l+4}
  +\frac{(-i\omega)^6}{384}\sum d_l^2d_{l+1}^3d_{l+3}\nonumber\\&&\strut
  +\frac{(-i\omega)^6}{768}\sum d_l^4d_{l+2}d_{l+3}
  +\frac{(-i\omega)^6}{96}\sum d_l^2d_{l+1}^3d_{l+2}
  +\frac{(-i\omega)^6}{192}\sum d_l^4d_{l+1}d_{l+2}\nonumber\\&&\strut
  +\frac{(-i\omega)^6}{128}\sum d_l^2d_{l+1}d_{l+2}^3
  +\frac{(-i\omega)^6}{1536}\sum d_l^4d_{l+1}^2
  +\frac{(-i\omega)^6}{64}\sum d_l^2d_{l+1}d_{l+2}\sum'd_{l'}^2\nonumber\\&&\strut
  +\frac{(-i\omega)^6}{128}\sum d_l^2d_{l+1}\sum'd_{l'}^2d_{l'+1}
  +\frac{(-i\omega)^6}{2304}\sum d_l^6\nonumber\\&&\strut
  +\frac{(-i\omega)^6}{256}\sum d_l^4\sum'd_{l'}^2
  +\frac{(-i\omega)^6}{384}\sum d_l^2\sum'd_{l'}^2\sum''d_{l''}^2+\ldots\bigg)
  \ =\label{sumord1}\\[12pt]
  &=&{\cal S}\exp\bigg(-i\omega d
  +\frac{(-i\omega)^2}4\sum d_l^2
  +\frac{(-i\omega)^3}8\sum d_l^2d_{l+1}
  +\frac{(-i\omega)^4}{16}\sum d_l^3d_{l+1}d_{l+2}\nonumber\\&&\strut
  +\frac{(-i\omega)^4}{64}\sum d_l^4
  +\frac{(-i\omega)^5}{32}\sum d_l^2d_{l+1}d_{l+2}d_{l+3}
  +\frac{(-i\omega)^5}{384}\sum d_l^4d_{l+2}\nonumber\\&&\strut
  +\frac{(-i\omega)^5}{64}\sum d_l^2d_{l+1}^3
  +\frac{(-i\omega)^5}{96}\sum d_l^4d_{l+1}
  +\frac{(-i\omega)^6}{64}\sum d_l^2d_{l+1}d_{l+2}d_{l+3}d_{l+4}\nonumber\\&&\strut
  +\frac{(-i\omega)^6}{384}\sum d_l^2d_{l+1}^3d_{l+3}
  +\frac{(-i\omega)^6}{768}\sum d_l^4d_{l+2}d_{l+3}
  +\frac{(-i\omega)^6}{96}\sum d_l^2d_{l+1}^3d_{l+2}\nonumber\\&&\strut
  +\frac{(-i\omega)^6}{192}\sum d_l^4d_{l+1}d_{l+2}
  +\frac{(-i\omega)^6}{128}\sum d_l^2d_{l+1}d_{l+2}^3
  +\frac{(-i\omega)^6}{1536}\sum d_l^4d_{l+1}^2\nonumber\\&&\strut
  +\frac{(-i\omega)^6}{2304}\sum d_l^6+\ldots\bigg).\label{sumord}
\end{eqnarray}
Here a prime at the summation symbol means disjunct indices, the symbol
${\cal S}$ indicates the summation ordering. Summation ordering defined here,
therefore, means that in a product of sums the indices are disjunct. As the
time ordering cannot be used directly for the exponent of a time ordered
product in quantum field theory, the same holds for the summation ordering: it
is defined only if the exponential series in Eq.~(\ref{sumord}) is expanded as
in Eq.~(\ref{sumord1}). However, the coefficients are all positive, and by
combinatorical means (worked out in Appendix~D), the coefficients of the
non-ordered series can be retained.

\section{Conclusion}
In this paper we have presented novel analytic results for an important class
of coupled map lattices (CMLs), so-called chaotic strings, which play an
important role in extensions of stochastically quantized field theories. We
were able to derive an analytic expression for the invariant 1-point density
of the coupled map lattice as a function of the coupling, extending previous
results obtained in Refs.~\cite{Groote:2006,Groote:2007}. The invariant
density was approximated to $p$-th order in an iterative recurrence scheme,
giving exact result for $p \to \infty$.

This scheme allowed us to better understand the nontrivial dependence of
expectations of observables (such as the vacuum expectation value) on the
coupling constant. A complex selfsimilar structure was found, for which we
developed a mathematical description in terms of scaling functions. Two types
of scaling functions were introduced (draft functions and blunt functions)
which allowed us to comprehensively describe topological changes, as well as
gradual changes, of the generating partition as a function of the coupling
$a$, thus understanding the corresponding changes of the integrals that occur
when expectation values are calculated.

Analytical expressions were found for both draft and the blunt functions.
Based on these, in the vicinity of local maxima the scale and parameter
invariance could be described in an analytic way by introducing a
renormalization group 'model' function (the analogue of Feigenbaum's fixed
point function for 1-dimensional critical maps). By implementing a graph
theoretical method, explicit calculations could be performed for this model
function, with surprising connections to Feynman graphs in quantum field
theory and other graphs (binary forests) developed to understand the
higher-order correlations of uncoupled Chebychev maps. The exponential draft
function as the main element of the model function was expressed as a
(summation ordered) exponential series.

Our results significantly advance the analytic understanding of chaotic
non-hyperbolic CMLs, of relevance in many different areas of physics,
and are in good agreement with direct numerical simulations of the CML.

\subsection*{Acknowledgements}
This work is supported in part by the Estonian target financed project
No.~0180056s09 and by the Estonian Science Foundation under grant No.~8769.
S.~Groote acknowledges support from a grant given by the Deutsche
Forschungsgemeinschaft for staying at Mainz University as guest scientist for
a couple of months. C.~Beck's research is supported by a Springboard
Fellowship of the Engineering and Physical Sciences Research Council (EPSRC).

\newpage

\begin{appendix}

\section{Explicit calculation for the first and second iterate}
\setcounter{equation}{0}\def\theequation{A\arabic{equation}}
In order to explain our method to obtain the first iterate in
Eq.~(\ref{rhoa1ext}), we calculate
\begin{eqnarray}
\lefteqn{\frac d{d\phi}T_{2a}^+\left(T_{2a}^+\left(\phi;
  T_2(\phi_+)+T_2(\phi_-)\right);\phi_++\phi_-\right)\ =}\nonumber\\
  &=&T_{2a}^{+\prime}\left(\phi;T_2(\phi_+)+T_2(\phi_-)\right)
  T_{2a}^{+\prime}\left(T_{2a}^+\left(\phi;T_2(\phi_+)+T_2(\phi_-)\right);
  \phi_++\phi_-\right).
\end{eqnarray}
Close to $\phi=1$ we obtain
\begin{eqnarray}
T_{2a}^{+\prime}(1-ax;T_2(\phi_+)+T_2(\phi_-))&=&\frac14+O(a),\nonumber\\
T_{2a}^+(1-ax;T_2(\phi_+)+T_2(\phi_-))&=&1-ax'+O(a^2)
\end{eqnarray}
where
\begin{equation}
x'=\frac x4+\frac18(T_2(\phi_+)+T_2(\phi_-)-2).
\end{equation}
Inserting the positive square root
$T_{2a}^+(1-ax;T_2(\phi_+)+T_2(\phi_-))\approx 1-ax'$ for $\phi'$ into
the expression $T_{2a}^{+\prime}(\phi';\phi_++\phi_-)$ does not lead to the
approximation~(\ref{approxp}). However, if we insert the negative square root
$T_{2a}^-(1-ax;T_2(\phi_+)+T_2(\phi_-))\approx-1+ax'$, we obtain
\begin{equation}
T_{2a}^{+\prime}(-1+ax';\phi_++\phi_-)
  \approx\frac1{2\sqrt{2a(1-a)}\sqrt{x/4-r_2^1(\phi_+)-r_2^1(\phi_-)}}
\end{equation}
and together with $T_{2a}^{-\prime}(1-ax;T_2(\phi_+)+T_2(\phi_-))\approx-1/4$
the integrand of the first iterate, which is now given by
\begin{equation}
-\frac d{d\phi}T_{2a}^+\left(T_{2a}^-\left(\phi;
  T_2(\phi_+)+T_2(\phi_-)\right);\phi_++\phi_-\right).
\end{equation}
For the second iterate, we use
\begin{eqnarray}
T_{2a}^{+\prime}(1-ax;T_2(T_2(\phi_+))+T_2(T_2(\phi_-)))&=&\frac14+O(a),
  \nonumber\\
T_{2a}^+(1-ax;T_2(T_2(\phi_+))+T_2(T_2(\phi_-)))&=&1-ax'+O(a^2),
  \nonumber\\
T_{2a}^{-\prime}(1-ax';T_2(\phi_+)+T_2(\phi_-))&=&-\frac14+O(a),\nonumber\\
T_{2a}^-(1-ax';T_2(\phi_+)+T_2(\phi_-))&=&-1+ax''+O(a^2)
\end{eqnarray}
with
\begin{eqnarray}
x'&=&\frac x4+\frac18(T_2(T_2(\phi_+))+T_2(T_2(\phi_-))-2),\nonumber\\
x''&=&\frac{x'}4+\frac18(T_2(\phi_+)+T_2(\phi_-)-2)
\end{eqnarray}
and, finally,
\begin{equation}
T_{2a}^{-1\prime}(-1+ax'';\phi_++\phi_-)\approx
  \frac1{2\sqrt{2a(1-a)}\sqrt{x/16-r_2^2(\phi_+)-r_2^2(\phi_-)}}.
\end{equation}
The integrand, therefore, is given by
\begin{equation}
-\frac d{d\phi}T_{2a}^+\left(T_{2a}^+\left(T_{2a}^-\left(\phi;
  T_2(T_2(\phi_+))+T_2(T_2(\phi_-))\right);T_2(\phi_+)+T_2(\phi_-)\right);
  \phi_++\phi_-\right).
\end{equation}

\section{Details on the ``path of the roots''}
\setcounter{equation}{0}\def\theequation{B\arabic{equation}}
In Appendix~A we have considered only one ``path of the roots'' which
directly leads to the iterates $\rho_a^{(p)}(\phi)$. However, it can be
shown that by adding over all possible paths, we end up with
Eq.~(\ref{pathroot}) which is equal to $\rho_a^{(0)}(\phi)$ close to the
boundary $\phi=-1$ and equal to the sum of the $\rho_a^{(q)}(\phi)$
(for $q\le p$) at the boundary $\phi=+1$. Looking at the region close to the
boundary $\phi=-1$, we expand the integrand of~(\ref{pathroot}) with
$\phi=ay-1$ for small values of $a$. Looking at different orders $p$, we
obtain
\begin{eqnarray}
\lefteqn{\frac d{d\phi}T_{2a}^{-1}(\phi;\phi_++\phi_-)
  \ :=\ \frac d{d\phi}T_{2a}^+(\phi;\phi_++\phi_-)
  +\frac d{d\phi}T_{2a}^-(\phi;\phi_++\phi_-)\ =}\nonumber\\
  &\approx&\frac1{\sqrt{16a(1-a)(y-r_2^0(\phi_+)-r_2^0(\phi_-))}}
  \ =\ \frac1{\sqrt{8a(1-a)(y-r_2^0(\phi_+)-r_2^0(\phi_-))}}\sin\pfrac\pi4,
  \nonumber\\
\lefteqn{\frac d{d\phi}T_{2a}^{-1}(T_{2a}^{-1}(\phi;T_2(\phi_+)+T_2(\phi_-));
  \phi_++\phi_-)
  \ =}\nonumber\\
  &:=&
  \frac d{d\phi}T_{2a}^+(T_{2a}^+(\phi;T_2(\phi_+)+T_2(\phi_-));\phi_++\phi_-)
  \nonumber\\&&\strut
  +\frac d{d\phi}T_{2a}^+(T_{2a}^-(\phi;T_2(\phi_+)+T_2(\phi_-));\phi_++\phi_-)
  \nonumber\\&&\strut
  +\frac d{d\phi}T_{2a}^-(T_{2a}^+(\phi;T_2(\phi_+)+T_2(\phi_-));\phi_++\phi_-)
  \nonumber\\&&\strut
  +\frac d{d\phi}T_{2a}^-(T_{2a}^-(\phi;T_2(\phi_+)+T_2(\phi_-));\phi_++\phi_-)
  \nonumber\\
  &\approx&\frac1{\sqrt{8a(1-a)(y-r_2^0(\phi_+)-r_2^0(\phi_-))}}\frac12\left(
  \frac1{\sqrt2\sqrt{2+\sqrt2}}+\frac1{\sqrt2\sqrt{2-\sqrt2}}\right)
  \nonumber\\
  &=&\frac1{\sqrt{8a(1-a)(y-r_2^0(\phi_+)-r_2^0(\phi_-))}}\frac12\left(
  \sin\pfrac\pi8+\sin\pfrac{3\pi}8\right),\nonumber\\
\lefteqn{\frac d{d\phi}T_{2a}^{-1}(T_{2a}^{-1}(T_{2a}^{-1}(\phi;
  T_2(T_2(\phi_+))+T_2(T_2(\phi_-)));T_2(\phi_+)+T_2(\phi_-));\phi_++\phi_-)
  \ =}\nonumber\\
  &:=&\frac d{d\phi}T_{2a}^+(T_{2a}^+(T_{2a}^+(\phi;
  T_2(T_2(\phi_+))+T_2(T_2(\phi_-)));T_2(\phi_+)+T_2(\phi_-));\phi_++\phi_-)
  \nonumber\\&&\strut
  +\frac d{d\phi}T_{2a}^+(T_{2a}^+(T_{2a}^-(\phi;
  T_2(T_2(\phi_+))+T_2(T_2(\phi_-)));T_2(\phi_+)+T_2(\phi_-));\phi_++\phi_-)
  \nonumber\\&&\strut
  +\frac d{d\phi}T_{2a}^+(T_{2a}^-(T_{2a}^+(\phi;
  T_2(T_2(\phi_+))+T_2(T_2(\phi_-)));T_2(\phi_+)+T_2(\phi_-));\phi_++\phi_-)
  \nonumber\\&&\strut
  +\frac d{d\phi}T_{2a}^+(T_{2a}^-(T_{2a}^-(\phi;
  T_2(T_2(\phi_+))+T_2(T_2(\phi_-)));T_2(\phi_+)+T_2(\phi_-));\phi_++\phi_-)
  \nonumber\\&&\strut
  +\frac d{d\phi}T_{2a}^-(T_{2a}^+(T_{2a}^+(\phi;
  T_2(T_2(\phi_+))+T_2(T_2(\phi_-)));T_2(\phi_+)+T_2(\phi_-));\phi_++\phi_-)
  \nonumber\\&&\strut
  +\frac d{d\phi}T_{2a}^-(T_{2a}^+(T_{2a}^-(\phi;
  T_2(T_2(\phi_+))+T_2(T_2(\phi_-)));T_2(\phi_+)+T_2(\phi_-));\phi_++\phi_-)
  \nonumber\\&&\strut
  +\frac d{d\phi}T_{2a}^-(T_{2a}^-(T_{2a}^+(\phi;
  T_2(T_2(\phi_+))+T_2(T_2(\phi_-)));T_2(\phi_+)+T_2(\phi_-));\phi_++\phi_-)
  \nonumber\\&&\strut
  +\frac d{d\phi}T_{2a}^-(T_{2a}^-(T_{2a}^-(\phi;
  T_2(T_2(\phi_+))+T_2(T_2(\phi_-)));T_2(\phi_+)+T_2(\phi_-));\phi_++\phi_-)
  \nonumber\\
  &\approx&\frac1{\sqrt{8a(1-a)(y-r_2^0(\phi_+)-r_2^0(\phi_-))}}
  \times\strut\nonumber\\&&\strut\times\frac14\Bigg(
  \frac1{\sqrt2\sqrt{2+\sqrt2}\sqrt{2+\sqrt{2+\sqrt2}}}
  +\frac1{\sqrt2\sqrt{2-\sqrt2}\sqrt{2+\sqrt{2-\sqrt2}}}\nonumber\\&&\strut
  +\frac1{\sqrt2\sqrt{2+\sqrt2}\sqrt{2-\sqrt{2+\sqrt2}}}
  +\frac1{\sqrt2\sqrt{2-\sqrt2}\sqrt{2-\sqrt{2-\sqrt2}}}\Bigg)\\
  &=&\frac1{\sqrt{8a(1-a)(y-r_2^0(\phi_+)-r_2^0(\phi_-))}}
  \frac14\left(\sin\pfrac\pi{16}+\sin\pfrac{3\pi}{16}
  +\sin\pfrac{7\pi}{16}+\sin\pfrac{5\pi}{16}\right).\nonumber
\end{eqnarray}
For obtaining the final results we have used that
\begin{eqnarray}
\frac1{\sqrt2}&=&\sin\pfrac\pi4,\nonumber\\
\frac1{\sqrt2\sqrt{2+\sqrt2}}&=&\frac12\sqrt{2-\sqrt2}
  \ =\ \sin\pfrac\pi8,\nonumber\\
\frac1{\sqrt2\sqrt{2-\sqrt2}}&=&\frac12\sqrt{2+\sqrt2}
  \ =\ \sin\pfrac{3\pi}8,\nonumber\\
\frac1{\sqrt2\sqrt{2+\sqrt2}\sqrt{2+\sqrt{2+\sqrt2}}}
  &=&\frac12\sqrt2\sqrt{2+\sqrt2}\sqrt{2-\sqrt{2+\sqrt2}}
  \ =\ \sin\pfrac\pi{16},\nonumber\\
\frac1{\sqrt2\sqrt{2-\sqrt2}\sqrt{2+\sqrt{2-\sqrt2}}}
  &=&\frac12\sqrt2\sqrt{2-\sqrt2}\sqrt{2-\sqrt{2-\sqrt2}}
  \ =\ \sin\pfrac{3\pi}{16},\nonumber\\
\frac1{\sqrt2\sqrt{2+\sqrt2}\sqrt{2-\sqrt{2+\sqrt2}}}
  &=&\frac12\sqrt2\sqrt{2+\sqrt2}\sqrt{2+\sqrt{2+\sqrt2}}
  \ =\ \sin\pfrac{7\pi}{16},\nonumber\\
\frac1{\sqrt2\sqrt{2-\sqrt2}\sqrt{2-\sqrt{2-\sqrt2}}}
  &=&\frac12\sqrt2\sqrt{2-\sqrt2}\sqrt{2+\sqrt{2-\sqrt2}}
  \ =\ \sin\pfrac{5\pi}{16}.
\end{eqnarray}
Because of
\begin{equation}
\frac1{2^p}\sum_{m=0}^{2^p-1}\sin\pfrac{\pi(2m+1)}{2^{p+2}}
  \to\int_0^1\sin\pfrac{\pi t}2dt=\frac2\pi,
\end{equation}
for higher and higher orders of $p$ the integrand approaches
\begin{equation}
\frac2\pi\ \frac1{\sqrt{8a(1-a)(y-r_2^0(\phi_+)-r_2^0(\phi_-))}}
  =\frac1{\pi\sqrt{2a(1-a)(y-r_2^0(\phi_+)-r_2^0(\phi_-))}}.
\end{equation}
Similar considerations lead to the iterates close to the boundary $\phi=+1$.

\section{The model function for $p\to\infty$}
\setcounter{equation}{0}\def\theequation{C\arabic{equation}}
Because the second part of the radicand in Eq.~(\ref{model1}) contains a
$k$-fold Chebychev map, one first has to unfold this map. Using
\begin{equation}
\phi_\pm=\sin\left(\frac\pi2t_\pm\right),\qquad
\rho_0(\phi_\pm)d\phi_\pm=\frac12dt_\pm,
\end{equation}
one obtains
\begin{eqnarray}
\lefteqn{\hat f^{(\hat p,k)}(\Delta\hat t;t_0)=\frac14\int_{-1}^1dt_+dt_-
  \times}\strut\\&&\strut\kern-7pt\times\cos\left(\sqrt{\pi^2\left(\Delta\hat t
  -2^{-\hat p}\hat a\ttilde d^{(k)}(t_0;t_+,t_-)\right)^2-2\hat ar_2^\infty
  \left(-\cos(2^{k-1}\pi t_+),-\cos(2^{k-1}\pi t_-)\right)}\right).\nonumber
\end{eqnarray}
After the substitution $t'_\pm=2^kt_\pm$ the result reads
\begin{eqnarray}
\lefteqn{\hat f^{(\hat p,k)}(\Delta\hat t;t_0)=\frac1{4^{k+1}}\int_{-2^k}^{2^k}
  dt'_+dt'_-\times}\strut\\&&\strut\kern-7pt\times
  \cos\left(\sqrt{\pi^2\left(\Delta\hat t-2^{-\hat p}\hat a\ttilde d^{(k)}(t_0;
  2^{-k}t'_+,2^{-k}t'_-)\right)^2-2\hat ar_2^\infty
  \left(-\cos(\frac\pi2t'_+),-\cos(\frac\pi2t'_-)\right)}\right).\nonumber
\end{eqnarray}
The huge integration range $[-2^k,2^k]$ for each of $\phi'_\pm$ can be divided
up into the standard intervals $[-1,+1]$ by using
\begin{equation}
t''_\pm=t'_\pm+2^k-2n_\pm-1\in[-1,+1],
\end{equation}
where $n_\pm=0,1,\ldots,2^k-1$ counts the intervals. Because of
$t_\pm=2^{-k}(t''_\pm+2n_\pm+1)-1$, the (twofold linearized) draft function
$\ttilde d^{(k)}(t_0;t_+,t_-)$ is nearly constant in each of the intervals
$[-1,+1]$. In defining
\begin{equation}
\hat d_{n_+n_-}(t_0)
  :=\lim_{k\to\infty}\ttilde d^{(k)}(t_0;2^{-k}(2n_++1)-1,2^{-k}(2n_-+1)-1)
\end{equation}
one has to deal only with the blunt. Using
\begin{eqnarray}
\lefteqn{-\cos(\frac\pi2t'_\pm)\ =\ -\cos\left(\frac\pi2t''_\pm
  -\frac\pi2(2^k-1-2n_\pm)\right)\ =}\nonumber\\
  &=&-\cos\left(\frac\pi2t''_\pm-2^{k-1}\pi+\frac\pi2+\pi n_\pm\right)
  \ =\ \sin\left(\frac\pi2t''_\pm+\pi n_\pm\right)\ =\nonumber\\
  &=&(-1)^{n_\pm}\sin(\frac\pi2t''_\pm)
\end{eqnarray}
and removing the sign by replacing $t''_\pm\to-t''_\pm$, for the limit
$k\to\infty$ one obtains
\begin{eqnarray}\label{modelint}
\hat f^{(\hat p)}(\Delta\hat t;t_0)&=&\lim_{k\to\infty}\frac1{4^{k+1}}
  \sum_{n_\pm=0}^{2^k-1}\int_{-1}^1dt''_+dt''_-
  \times\strut\nonumber\\&&\strut\kern-14pt\times
  \cos\left(\sqrt{\pi^2\left(\Delta\hat t
  -2^{-\hat p}\hat a\hat d_{n_+n_-}(t_0)\right)^2
  -2\hat ar_2^\infty\left(\sin(\frac\pi2t''_+),\sin(\frac\pi2t''_-)\right)}
  \right)\ =\nonumber\\
  &=&\lim_{k\to\infty}\frac1{4^k}\sum_{n_\pm=0}^{2^k-1}\int\rho_0(\phi_+)
  d\phi_+\rho_0(\phi_-)d\phi_-\times\strut\nonumber\\&&\strut\times
  \cos\left(\sqrt{\pi^2\left(\Delta\hat t
  -2^{-\hat p}\hat a\hat d_{n_+n_-}(t_0)\right)^2
  -2\hat ar_2^\infty\left(\phi_+,\phi_-\right)}\right).
\end{eqnarray}
In the final step, $\phi_\pm=\sin(\pi t''_\pm/2)$ is used. The integrand does
depend on $\phi_\pm$ only via $r_2^\infty(\phi_+,\phi_-)$ while for the first
part one can use the abbreviation
\begin{equation}
\hat c=\pi\left(\Delta\hat t-2^{-\hat p}\hat a\hat d_{n_+n_-}(t_0)\right).
\end{equation}
\begin{figure}\begin{center}
\epsfig{figure=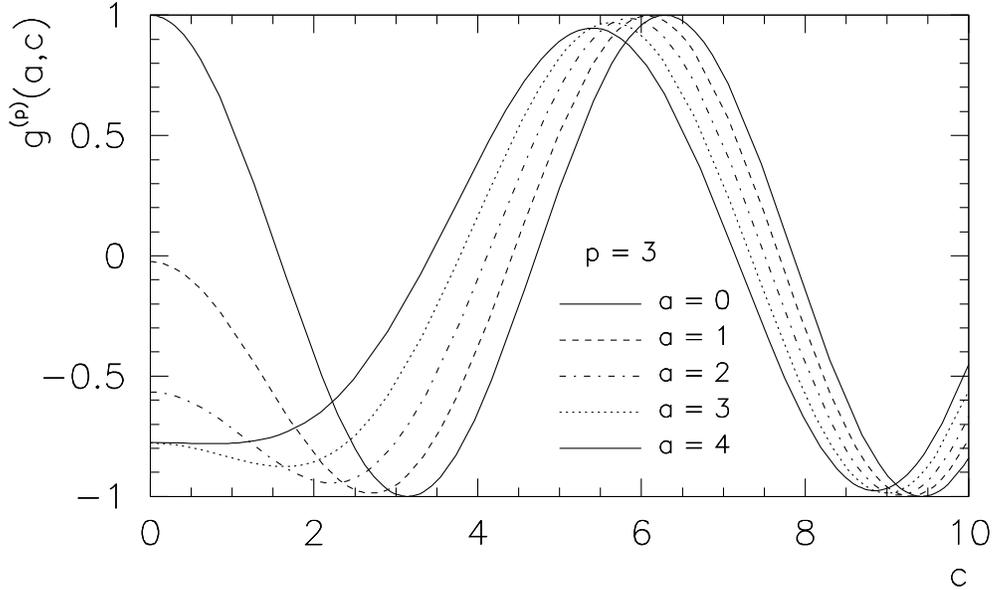, scale=0.8}
\caption{\label{inteles}$g^{(p)}(\hat a,\hat c)$ as a function of $\hat c$
for different values of $\hat a$. The order $p$ of the blunt function
$r_2^p(\phi_+,\phi_-)$ is chosen to be $p=3$.}
\end{center}
\end{figure}
In Fig.~\ref{inteles} the function
\begin{equation}
g^{(p)}(\hat a,\hat c):=\int\rho_0(\phi_+)d\phi_+\rho_0(\phi_-)d\phi_-
  \cos\left(\sqrt{\hat c^2-2\hat ar_2^p(\phi_+,\phi_-)}\right)
\end{equation}
is shown as a function of $\hat c$ for different values of $\hat a$. For the
order $p$ of the blunt function $r_2^p(\phi_+,\phi_-)$ a value of $p=3$ is
enough to provide sufficient precision. In analysing the position of the
drafted minima in dependence on $\hat a$, one obtains\footnote{The value
$-8/3$ is not rigorously derived but follows from numerical inspection.}
\begin{equation}
\hat c(\hat a)^2-\hat c(0)^2=-\frac{8\hat a}3\quad\Leftrightarrow\quad
\hat c(0)=\sqrt{\hat c(\hat a)^2+\frac{8\hat a}3},
\end{equation}
leading to
\begin{equation}
g^{(p)}(\hat a,\hat c)=\cos\left(\sqrt{\hat c^2+\frac{8\hat a}3}\right).
\end{equation}
This function can be inserted into Eq.~(\ref{modelint}) to give
\begin{equation}
\hat f^{(\hat p)}(\Delta\hat t;t_0)=\lim_{k\to\infty}\frac1{4^k}
  \sum_{n_\pm=0}^{2^k-1}\cos\left(\sqrt{\pi^2\left(\Delta\hat t
  -2^{-\hat p}\hat a\hat d_{n_+n_-}(t_0)\right)+\frac{8\hat a}3}\right).
\end{equation}
In the limit $k\to\infty$, the remaining summations over $n_\pm$ can be
reunited to an integration over $\phi_\pm$, leading to the final result in
Eq.~(\ref{model2}). For practical reasons, however, $k$ is kept finite in the
main text.

\section{Combinatorics for the exponential draft function}
\setcounter{equation}{0}\def\theequation{D\arabic{equation}}
The (positive) coefficients of the exponent of the summation ordered
exponential draft function can be calculated by using combinatorial
considerations. In this appendix we demonstrate the procedure for single and
multiple trees in the summation ordered expression and show how to derive
the coefficients of the non-ordered expression.

\subsection{Coefficients for single trees}
According to our graph theory, single binary trees represent unique relations
between powers of $2$. The sum $\sum d_l^2d_{l+1}^3d_{l+3}$ for instance,
represented by the single tree
\begin{center}
\epsfig{figure=tree23010.eps, scale=0.5}
\end{center}
results from the unique relation $1+1+2+2+2-8=0$. In general, from
combinatorics one can see that for the expression
\begin{equation}
\sum_{l=0}^{k-n+1}\prod_{i=0}^{n-2}d_{l+i}^{n_i},\qquad
\sum_{i=0}^{n-2}n_i=n
\end{equation}
there are $n!/(n_0!n_1!\cdots n_{n-2}!)$ discriminable permutations without
taking into account the global sign (which gives a factor $2!$). Together with
a factor $1/n!$ for the exponential series in Eq.~(\ref{edraft}) and a factor
$1/2^n$ from the trigonometric functions $T_{2^l}(\sin(\pi t_+/2))$ after
their expansion in exponential functions, one obtains
\begin{equation}
\frac1{n!2^n}\pfrac{2!n!}{n_0!n_1!\cdots n_{n-2}!}
  =\frac1{2^{n-1}n_0!n_1!\cdots n_{n-2}!}.
\end{equation}

\subsection{Coefficients for multiple trees}
While the coefficients of the single tree contributions are the same for
the summation ordered as for the non-ordered exponential draft function, the
coefficients of the multiple tree contributions are different in both cases.
However, the coefficients of the summation ordered contributions can be
calculated by combinatorial means. This is shown for the contributions of
orders $n=4$ to $6$ in the following. Multiple binary trees correspond to
relations between powers of $2$ which can be split into two or more single
relations separated by a vertical bar.
\begin{itemize}
\item$\sum d_l^4$: $+1-1|+1-1=0$ gives $4!/2!/2!$ permutation if $+1$ and $-1$
are taken to be two different elements. A global sign for the first or the
second part $+1-1=0$ need not be taken into account because the permutations
already lead to the sign-changed contributions (sign non-sensitive). Together
with the global factor $1/4!/2^4$ one obtains
\[\frac1{4!2^4}\pfrac{4!}{2!2!}=\frac1{2^42!2!}=\frac1{64}.\]
\item$\sum d_l^4d_{l+1}$: $+1+1-2|+1-1=0$ gives $5!/3!1!1!$ permutations where
now the global sign for the first part has to be taken into account. One
obtains two (formally equal) contributions
\[\frac1{5!2^5}\left(\frac{5!}{3!1!1!}+\frac{5!}{1!3!1!}\right)
  =\frac2{2^53!}=\frac1{96}.\]
\item$\sum d_l^2d_{l+1}^3$: $+1+1-2|+2-2=0$ gives $5!/2!1!2!$ permutations
where the same holds as in the previous case. One obtains
\[\frac1{5!2^5}\left(\frac{5!}{2!1!2!}+\frac{5!}{2!2!1!}\right)
  =\frac2{2^52!2!}=\frac1{128}.\]
\item$\sum d_l^2d_{l+1}^3d_{l+2}$: $+1+1-2|+2+2-4=0$ gives $6!/2!1!2!1!$
permutations. In this case one has to take into account the global signs of
both parts. In addition the numer of elements changes if the global signs
change. One obtains
\[\frac1{6!2^6}\left(\frac{6!}{2!1!2!1!}+\frac{6!}{2!3!1!}+\frac{6!}{2!1!2!1!}
  +\frac{6!}{2!3!1!}\right)=\frac1{96}.\]
\item$\sum d_l^4d_{l+1}d_{l+2}$: $+1-1|+1+1+2-4=0$ gives
\[\frac1{6!2^6}\left(\frac{6!}{3!1!1!1!}+\frac{6!}{1!3!1!1!}\right)
  =\frac1{192}.\]
\item$\sum d_l^2d_{l+1}d_{l+2}^3$: $+1+1+2-4|+4-4=0$ gives
\[\frac1{6!2^6}\left(\frac{6!}{2!1!1!2!}+\frac{6!}{2!1!2!1!}\right)
  =\frac1{128}.\]
\item$\sum d_l^4d_{l+1}^2$: Besides the sign sensitive possibility
$+1+1-2|+1+1-2=0$, in principle there is also a non-sensitive possibility with
three parts, $+1-1|+1-1|+2-2=0$. However, this possibility is the same as the
sign-changed possibility $+1+1-2|-1-1+2=0$. Therefore, only the first has to
be taken into account, resulting in
\[\frac1{6!2^6}\left(\frac{6!}{4!2!}+\frac{6!}{2!2!1!1!}+\frac{6!}{2!2!1!1!}
  +\frac{6!}{4!2!}\right)=\frac1{1536}.\]
\item$\sum d_l^6$: $+1-1|+1-1|+1-1=0$ contains three parts not sensitive to
the global signs. Therefore, one obtains
\[\frac1{6!2^6}\pfrac{6!}{3!3!}=\frac1{2304}.\]
\end{itemize}

\subsection{Summation ordered and non-ordered coefficients}
The relation between the coefficients of the summation ordered and
non-ordered exponential draft function can be found by using
\begin{eqnarray}\label{double}
\lefteqn{\sum d_l^2\sum d_{l'}^2\ =\ \sum d_l^2\sum'd_{l'}^2+\sum d_l^4,}
  \nonumber\\[7pt]
\lefteqn{\sum d_l^2d_{l+1}\sum d_{l'}^2\ =\ \sum d_l^2d_{l+1}\sum'd_{l'}^2
  +\sum d_l^4d_{l+1}+\sum d_l^2d_{l+1}^3,}\nonumber\\[12pt]
\lefteqn{\sum d_l^2d_{l+1}d_{l+2}\sum d_{l'}^2
  \ =\ \sum d_l^2d_{l+1}d_{l+2}\sum'd_{l'}^2
  +\sum d_l^4d_{l+1}d_{l+2}+\sum d_l^2d_{l+1}^3d_{l+2}
  +\sum d_l^2d_{l+1}d_{l+2}^3,}\nonumber\\[12pt]
\lefteqn{\sum d_l^2d_{l+1}\sum d_{l'}^2d_{l'+1}\ =}\nonumber\\
  &=&\sum d_l^2d_{l+1}\sum'd_{l'}^2d_{l'+1}
  +\sum d_l^2d_{l+1}^3d_{l+2}+\sum d_l^4d_{l+1}^2
  +\sum d_{l'}^2d_{l'+1}^3d_{l'+2}\ =\nonumber\\
  &=&\sum d_l^2d_{l+1}\sum'd_{l'}^2d_{l'+1}
  +\sum d_l^4d_{l+1}^2+2\sum d_l^2d_{l+1}^3d_{l+2},\nonumber\\[7pt]
\lefteqn{\sum d_l^4\sum d_{l'}^2\ =\ \sum d_l^4\sum'd_{l'}^2+\sum d_l^6,}
  \nonumber\\[7pt]
\lefteqn{\sum d_l^2\sum d_{l'}^2\sum d_{l''}^2
  \ =\ \Big(\sum d_l^2\sum'd_{l'}^2\Big)\sum d_{l''}^2
  +\sum d_l^4\sum d_{l''}^2\ =}\nonumber\\
  &=&\sum d_l^2\sum'd_{l'}^2\sum''d_{l''}^2+\sum d_l^2\sum'd_{l'}^4
  +\sum d_l^4\sum'd_{l'}^2+\sum d_{l'}^4\sum'd_{l''}^2+\sum d_l^6\ =
  \qquad\qquad\ \nonumber\\
  &=&\sum d_l^2\sum'd_{l'}^2\sum''d_{l''}^2+3\sum d_l^4\sum'd_{l'}^2
  +\sum d_l^6.
\end{eqnarray}
Using, for instance, for the contribution of order $n=6$ in
Eq.~(\ref{expseries}) a general ansatz and applying Eqs.~(\ref{double}), one
obtains
\begin{eqnarray}
\lefteqn{A_1\sum d_l^2d_{l+1}d_{l+2}d_{l+3}d_{l+4}
  +A_2\sum d_l^2d_{l+1}^3d_{l+3}+A_3\sum d_l^4d_{l+2}d_{l_3}}
  \nonumber\\[7pt]&&\strut
  +A_4\sum d_l^2d_{l+1}^3d_{l+2}+A_5\sum d_l^4d_{l+1}d_{l_2}
  +A_6\sum d_l^2d_{l+1}d_{l+2}^2+A_7\sum d_l^4d_{l+1}^2\nonumber\\[7pt]&&\strut
  +A_8\sum d_l^2d_{l+1}d_{l+2}\sum d_{l'}^2
  +A_9\sum d_l^2d_{l+1}\sum d_{l'}^2d_{l'+1}+A_{10}\sum d_l^6\nonumber\\[7pt]&&\strut
  +A_{11}\sum d_l^4\sum d_{l'}^2+A_{12}\sum d_l^2\sum d_{l'}^2\sum d_{l''}^2
  \ =\nonumber\\[7pt]
  &=&A_1\sum d_l^2d_{l+1}d_{l+2}d_{l+3}d_{l+4}
  +A_2\sum d_l^2d_{l+1}^3d_{l+3}+A_3\sum d_l^4d_{l+2}d_{l_3}\nonumber\\[7pt]&&\strut
  +(A_4+A_8+2A_9)\sum d_l^2d_{l+1}^3d_{l+2}
  +(A_5+A_8)\sum d_l^4d_{l+1}d_{l_2}\nonumber\\&&\strut
  +(A_6+A_8)\sum d_l^2d_{l+1}d_{l+2}^2+A_7\sum d_l^4d_{l+1}^2
  +A_8\sum d_l^2d_{l+1}d_{l+2}\sum'd_{l'}^2\nonumber\\&&\strut
  +A_9\sum d_l^2d_{l+1}\sum'd_{l'}^2d_{l'+1}
  +(A_{10}+A_{11}+A_{12})\sum d_l^6\nonumber\\&&\strut
  +(A_{11}+3A_{12})\sum d_l^4\sum'd_{l'}^2
  +A_{12}\sum d_l^2\sum'd_{l'}^2\sum''d_{l''}^2.
\end{eqnarray}
Comparing with Eq.~(\ref{sumord1}) one obtains
\begin{equation}
A_1=\frac1{64},\quad A_2=\frac1{384},\quad A_3=\frac1{768},\quad
A_8=\frac1{64},\quad A_9=\frac1{128},\quad A_{12}=\frac1{384},
\end{equation}
furthermore
\[A_{11}=\frac1{256}-3A_{12}=\frac1{256}-\frac1{128}=-\frac1{256},\]
and finally the coefficients for \ldots
\begin{eqnarray}
\sum d_l^2d_{l+1}^3d_{l+2}:&&A_4=\frac1{96}-A_8-2A_9
  =\frac1{96}-\frac1{64}-2\frac1{128}=-\frac1{48},\nonumber\\
\sum d_l^4d_{l+1}d_{l+2}:&&A_5=\frac1{192}-A_8
  =\frac1{192}-\frac1{64}=-\frac1{96},\nonumber\\
\sum d_l^2d_{l+1}d_{l+2}^3:&&A_6=\frac1{128}-A_8
  =\frac1{128}-\frac1{64}=-\frac1{128},\nonumber\\
\sum d_l^4d_{l+1}^2:&&A_7=\frac1{1536}-A_9
  =\frac1{1536}-\frac1{128}=-\frac{11}{1536},\nonumber\\
\sum d_l^6:&&A_{10}=\frac1{2304}-A_{11}-A_{12}
  =\frac1{2304}+\frac1{256}-\frac1{384}=\frac1{576}.\qquad
\end{eqnarray}

\end{appendix}

\end{document}